\documentclass[aps,prx,twocolumn,groupedaddress]{revtex4-2}
\usepackage{graphicx}
\usepackage{amsmath}
\usepackage{siunitx}

\usepackage{color}
\usepackage{ulem}

\begin{document}


\title{Measuring the magnetic dipole transition of single nanorods by {spectroscopy and} Fourier microscopy}


\author{Reinaldo Chacon$^{1}$}
\author{Aymeric Leray$^{1}$}
\author{Jeongmo Kim$^{2}$}
\author{Khalid Lahlil$^{2}$}
\author{Sanro Mathew$^{1}$} 
\author{Alexandre Bouhelier$^{1}$}
\author{Jong-Wook Kim$^{2}$}
\author{Thierry Gacoin$^{2}$}
\author{G\'erard {Colas des Francs$^{1}$}}\email{gerard.colas-des-francs@u-bourgogne.fr}
\affiliation{$^{1}$ Laboratoire Interdisciplinaire Carnot de Bourgogne (ICB), UMR 6303 CNRS, Universit\'e Bourgogne Franche-Comt\'e,
9 Avenue Savary, BP 47870, 21078 Dijon Cedex, France}

\affiliation{$^{2}$Physique de la Mati\`ere Condens\'ee, CNRS UMR 7643, Ecole Polytechnique, 91128 Palaiseau, France}


\date{\today}

\begin{abstract}
Rare-earth doped nanocrystals possess optical transitions with significant either electric or magnetic dipole characters.  They are of strong interest for understanding and engineering light-matter interactions at the nanoscale with numerous applications in nanophotonics. Here, we study the $^5$D$_0\rightarrow ^7$F$_1$ transition dipole vector in individual NaYF$_4$:Eu$^{3+}$ nanorod crystals by Fourier and confocal microscopies. {Single crystalline host matrix leads to narrow emission lines at room temperature that permit to separate Stark sublevels resulting from the crystal field splitting}. We observe a fully magnetic transition and {low variability} of the transition dipole orientation over several single nanorods. We estimate the proportion of the dipole transitions for the Stark sublevels.  We also determine an effective altitude of the rod with respect to the substrate. The narrow emission lines characteristic of NaYF$_4$:Eu$^{3+}$ ensure well-defined electric or magnetic transitions, and are thus instrumental for probing locally their electromagnetic environment by standard confocal microscopy.
\end{abstract}


\maketitle

\section{Introduction}
Light-matter interaction in the optical regime is principally of electric nature because the magnetic contribution is orders of magnitude weaker. Therefore, engineering optical magnetic dipole (MD) opens novel design rules for developing metamaterials \cite{Shalaev:07} and optical antennas \cite{Bidault-Mivelle-Bonod:19}.  Examples are artificial MDs created by specific ring-like geometries \cite{Devaux-Dereux-Bourillot-Weeber-Lacroute-Goudonnet-Girard:2000,Mary-Dereux-Ferrell:2005,Shafei:13,leFeber-Kuipers:13}. Of particular interest in this context are rare-earth ions because they naturally have MD optical transitions \cite{Taminiau-Zia:2012} that can be manipulated by the crystalline or molecular hosting environment.

Rare-earth doped crystals have original spectroscopic properties and extensive theoretical and experimental efforts have been devoted to investigate their distinctive luminescence features. They are widely produced for solid state lasers or as phosphors for lightening \cite{Blasse-Grabmaier:94}, but several other promising applications are also envisioned.  Since some emission lines are very sensitive to the local temperature, rare-earth doped nanocrystals have been used to measure the internal temperature of tumor cells during thermally induced apoptosis \cite{vetrone:10} or used as temperature sensor scanned above an electronic circuit \cite{Saidi:2009}. Luminescent nanocrystals are also considered as active probes for scanning near-field optical microscopy delivering a signal proportional to the local electric field \cite{Aigouy-deWilde-Mortier:2003}. Moreover, some transitions present a magnetic dipole activity used for probing the magnetic near-field \cite{Kasperczyk:15,Wiecha-Cuche:19}. Finally lifetime measurements of the electric dipole (ED) and MD allowed transitions in rare-earth doped nanocrystal were implemented to experimentally measure the local density of electromagnetic states (LDOS) \cite{Karaveli-Zia:2011,Aigouy:14,Rabouw-Norris:16,Li-Zia:18}. As a consequence, these luminescent nanoparticles constitute a family of nanoprobes extremely versatile and a complete understanding and characterization of these emission lines are a prerequisite to unlock the full potential of applications. 

In this work, we are studying a specific transition of NaYF$_4$:Eu$^{3+}$ nanocrystals. The spectrum of trivalent europium ions strongly depends on the site symmetry and are thus sensitive to the electronic structure. For instance, the $^5$D$_0\rightarrow ^7$F$_0$ transition is related to the coordination number of the Eu$^{3+}$ and the intensity of the robustness of the $^5$D$_0\rightarrow ^7$F$_1$ transition may be used as a reference for spectroscopic calibration \cite{Binnemans:15}.  This $^5$D$_0\rightarrow ^7$F$_1$ transition has a mainly MD character \cite{Binnemans:15}. In 1941, Freed and Weissman analyzed wide angle emission and concluded to the multipole nature of this line for randomly oriented emitters in solution \cite{Freed-Weissman:1941}. In 1980, Kunz and Lukosz measured the lifetime in thin films of Eu$^{3+}$ doped benzoyltrifluoroacetone chelate and quantified the magnetic/electric ratio of the transition and found a proportion of $80\%$ MD and $20\%$ ED in the spectral range $\lambda=592 \pm 5$ nm \cite{Kunz-Lukosz:80}. Last, in 2012 Taminiau {\it et al.} adapted the method of Freed and Weissman by using Fourier microscopy and obtained that the transition is $91\pm2 \%$ MD for Y$_2$O$_3$:Eu$^{3+}$ doped thin film \cite{Taminiau-Zia:2012}. All these samples involve randomly oriented emitters but in rare-earth doped nanocrystals, the single crystalline host matrix leads to ED or MD optical transitions that exhibit strong polarization properties directly linked to the crystal orientation \cite{Sayre-Freed:1956,Kim-Gacoin:17}. The dipole transition has a fixed orientation with respect to the rod axis and constitutes a {\it vectorial} nanoprobe.  Also of strong interest here, we use a single crystalline NaYF$_4$ host matrix for the Eu$^{3+}$ ions, limiting multiphonon relaxation channels and ensuring well separated transition lines \cite{Rabouw-Norris:16}. Indeed, the $^5$D$_1\rightarrow ^7$F$_3$ is an ED transition with an emission wavelength ($\lambda=585$ nm) close to the $^5$D$_0\rightarrow ^7$F$_1$ MD transition ($\lambda=592$ nm). Narrow lines are thus needed to carefully characterize the multipole nature of the transitions. The recorded data are correctly interpreted by considering a fully MD transition. We measure the MD orientation and quantify the contribution of MD transitions towards the Stark sublevels resulting from the crystal field splitting of the otherwise degenerated $^7$F$_1$ level. The work is organized as follows. In Sec. \ref{sect:dipole}, we show the experimental measurements and modeling of the spectroscopic properties of individual europium doped single crystalline NaYF$_4$ nanorods. We record spectrally-resolved polarized emission diagram and deduce the nature of the $^5$D$_0\rightarrow ^7$F$_1$ transition as well as its dipole moment orientation. In Sec. \ref{sect:Fourier}, we measure the rod emission by Fourier plane microscopy and discuss the contribution of the MDs within the spectral range. 

\section{Optical transition  dipole vector}
\label{sect:dipole}
\subsection{Electric and magnetic dipole optical transitions}
\label{sect:ED-MD}
\begin{figure}[h!]
 	\includegraphics[width=8cm]{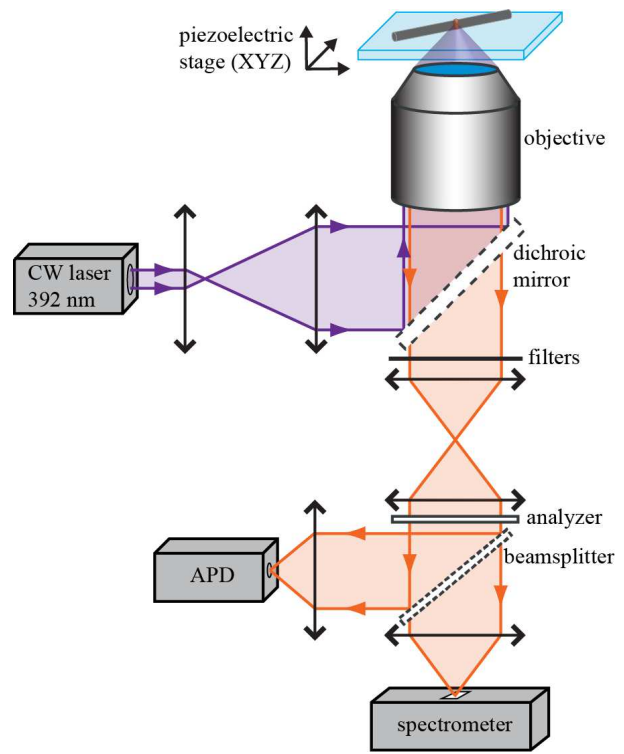}
	\caption{Schematic representation of the confocal optical microscope. {The Europium ions are excited by a continuous-wave laser emitting at a wavelength of 392 nm and focused by a microscope objective (40$\times$,  NA=0.6 or 60$\times$  NA=1.49, Nikon). The luminescence is collected by the same objective and is separated from the excitation laser by a dichroic mirror and routed either to an avalanche photodiode (APD) for confocal imaging or to a spectrometer. The polarisation of the emitted light is analyzed placing a polarizer after the sample ("analyzer").} 
	\label{fig:confocal}}
\end{figure}
\begin{figure}[h!]
	\includegraphics[width=6cm]{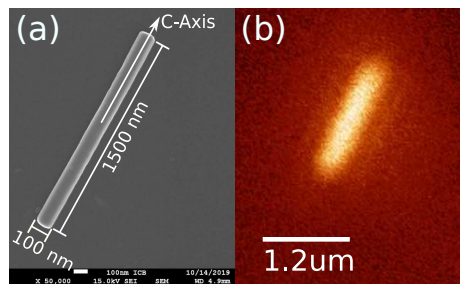}
	\caption{(a) SEM image of a NaYF$_4$:Eu$^{3+}$ single nanorod with a diameter of 90 nm and a length of 1500 nm. (b) Confocal image of the luminescence emitted by the {\it same} rod. 
		\label{fig:DipoleAngle}}
\end{figure}
\begin{figure}[h!]
	\includegraphics[width=8cm]{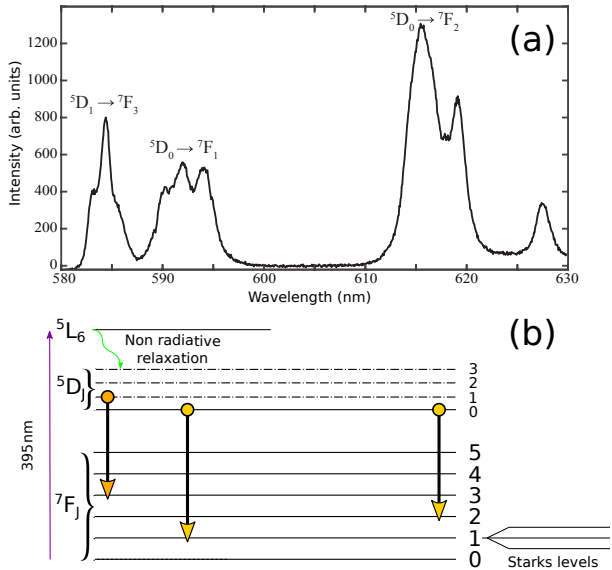}
	\caption{(a) Spectrum of a single NaYF$_4$:Eu$^{3+}$  nanorod (obtained without the analyzer, and with an {objective NA=0.6}). (b) Energy diagram of trivalent Europium. The ion has a $4f^6$ configuration with optical transitions occurring between the levels $^5D_1$ or $^5D_0$ and $^7F_J$. Each level has a degeneracy of $2J+1$ split by the crystalline field into different Stark levels.
\label{fig:EDiagramSpectra}}
\end{figure}

Europium doped single crystalline NaYF$_4$:Eu$^{3+}$ nanorods with an average diameter of 90 nm and a length of 1500 nm are synthesized as described in Ref. \cite{Wang-Liu:2010,FickOptExp:18} and are randomly dispersed on a quartz substrate (see appendix \ref{sect:Synthesis}). The emission of single nanorod is acquired with a home made epi-confocal microscope, depicted in Fig. \ref{fig:confocal}. Depending on the measurement, we use either a low numerical objective (NA = 0.6) or a high NA lens (NA =1.49).  A power of $\SI{236}{\micro \watt}$ at the focal point is necessary to generate a detectable luminescence signal. To be sure that emission spectra are originating from nanorods, a confocal image of the luminescence  is first acquired by scanning the sample with a piezoelectric stage and collecting the light with a fiber coupled APD, see Fig. \ref{fig:DipoleAngle}(b). The center of the nanorod is then positioned to the focal area and an emission spectrum is recorded, see Fig. \ref{fig:EDiagramSpectra}(a). Scanning electron micrographs (SEM) of the measured rods are acquired {\it a posteriori} and only single rod measurements are kept in the analysis, see Fig. \ref{fig:DipoleAngle}.

The photoluminescence spectra of a single rod is characterized by distinct emission lines, see Figure \ref{fig:EDiagramSpectra}(a) corresponding to transitions from level $^5D_0$ or $^5D_1$ to $^7F_J$ with $J=$0\ldots 6. Each transition has a degeneracy of $2J+1$, that can be lifted due to the crystal field perturbation splitting of the $^7F_J$ levels into different Stark sublevels [Figure \ref{fig:EDiagramSpectra}(b)] \cite{Binnemans:15}. According to the Judd-Ofelt model \cite{WalshBM}, all these transitions correspond to ED except for the $^5D_0\rightarrow ^7F_ 1$ transition (peak at 592 nm), which behaves as a MD. Nanorods exhibit the hexagonal $\beta$-NaYF$_4$ structure with the c-axis corresponding to the main axis of the rods. The symmetry group of the Y site is C$_{3h}$. We thus expect two emission lines for the $^5D_0 \rightarrow ^7$F$_1$ transition and one emission line for the $^5D_0 \rightarrow ^7$F$_2$ transition \cite{Binnemans:15}. However, because of the larger ionic radius of Eu$^{3+}$ compare to Y$^{3+}$, a symmetry breaking down to a lower C$_s$ symmetry occurs in presence of doping Eu$^{3+}$ at Y sites \cite{Tu-Chen:13}. Therefore, crystal-field splitting lifts the degeneracy of the Stark sublevels and we observe three emission lines for the  $^5D_0 \rightarrow ^7$F$_1$ MD transition but resolve three of the five emission lines for the  $^5D_0 \rightarrow ^7$F$_2$ ED transition. In the following, we use confocal microscopy to study the $^5D_0 \rightarrow ^7$F$_1$ MDs orientation within individual rods. We  measure the orientation of the dipole transition that presents a fixed angle with respect to the c-axis.

\subsection{Polarized emission diagram}
\subsubsection{Model of the luminescence emitted by rare-earth doped nanorods}   
\label{sect:Model}
\begin{figure}[h!]
		\includegraphics[width=6cm]{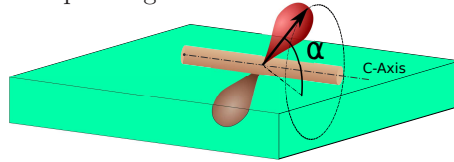}
	\caption{Orientation of the dipole transition moment for a rod on the substrate. The emission is modeled by dipolar emitters oriented at a fixed angle with respect to the c-axis (cone of semi-angle $\alpha$). 
		\label{fig:RodDipole}}
\end{figure}

\paragraph{Nanocrystal emission}
We consider a single rod with its c-axis aligned with the substrate/air interface (the general case is derived in the appendix \ref{sect:caxisBeta}). The ED or MD moments associated to a transition line have a fixed angle $\alpha$ with respect to the c-axis as illustrated in Fig. \ref{fig:RodDipole}. We assume a MD in the following, but similar expressions hold for the electric field emitted by an ED (see appendix \ref{sect:Green}). {Due to random distribution of emitters among the Y host sites of the matrix, the rod emission is modeled by the incoherent emission of dipolar emitters presenting a fixed orientation with respect to the c-axis (angle $\alpha$), leaving free the angle $\psi$ in the plane perpendicular to the c-axis.  The dipole vector is ${\bf p}=p_0(\cos \alpha,\sin \alpha \sin \psi,\sin \alpha \cos \psi)$ with $\psi \in [0; 2\pi[$.}
The magnetic field scattered at ${\bf r}$ by a dipole located at ${\bf r_0}$ can be expressed thanks to the {magnetic Green's tensor ${\bf  Q}$}
\begin{eqnarray} 
{\bf H}({\bf r})=\mu_0\omega_0^2 {\bf  Q}({\bf r},{\bf r_0},\omega_0) \cdot {\bf p}
\end{eqnarray}
where $\omega_0$ refers to the angular frequency of the oscillating dipole, corresponding to the emission line. The detected intensity in the medium of optical index $n_3$ follows $I({\bf r})=\left \vert {\bf H}({\bf r})\right \vert ^2/n_3$. Finally, for a random orientation $\psi$ in the plane perpendicular to the c-axis, the detected intensity is the incoherent emission of all the emitting dipoles so that it can be expressed as (symbol $\propto$ means proportional to)
\begin{widetext}
\begin{eqnarray} 
{\bf I}({\bf r})&=&\frac{1}{2\pi n_3}\int_0^{2\pi}\left \vert {\bf H}({\bf r})\right \vert ^2 d\psi
\propto  \left( \left \vert Q_{xx}\right \vert ^2+\left \vert Q_{yx}\right \vert ^2+\left \vert Q_{zx}\right \vert ^2\right)\cos^2\alpha
\\
\nonumber
&&
+ \left( \left \vert Q_{xy}\right \vert ^2+\left \vert Q_{yy}\right \vert ^2+\left \vert Q_{zy}\right \vert ^2\right) \frac{1}{2}\sin^2\alpha 
+ \left( \left \vert Q_{xz}\right \vert ^2+\left \vert Q_{yz}\right \vert ^2+\left \vert Q_{zz}\right \vert ^2\right)  \frac{1}{2}\sin^2\alpha 
\end{eqnarray}
\end{widetext}
that is equivalent to the incoherent emission of three orthogonal dipoles 
\begin{eqnarray}
\nonumber
{\bf p}_1&=&
\begin{pmatrix}
\cos\alpha \\0 \\ 0
\end{pmatrix}\,,
{\bf p}_2=\frac{1}{\sqrt{2}}
\begin{pmatrix}
0 \\ \sin \alpha  \\0 
\end{pmatrix} \,,
{\bf p}_3=\frac{1}{\sqrt{2}}
\begin{pmatrix}
0 \\0 \\ \sin\alpha
\end{pmatrix} \,.
\end{eqnarray}

\paragraph{Paraxial approximation}
For a detector located on the confocal optical axis, just after an analyzer, the asymptotic expansion of the Green's tensor leads to  the polarized emission
(see appendix \ref{sect:Green} for details):  
\begin{eqnarray}
I(\Phi_A)=A \sin^2\Phi_A +B \cos^2\Phi_A
\label{eq:paraxial}
\end{eqnarray}
where $\Phi_A$ refers to the angle of the analyzer with respect to the rod c-axis. $A=(\sin^2 \alpha)/2, B=\cos^2 \alpha$ for an electric dipole, and  $A=\cos^2 \alpha, B=(\sin^2 \alpha)/2$ for a magnetic dipole. So, the emitter angle with respect to the c-axis can be estimated from the fitting parameters $A,B$ of the measured polarized diagrams discussed later in \S \ref{sect:lowNA}.  Numerical simulations (not shown) reveal that this expression also correctly reproduces  the polarized diagram detected with a low NA objective. Note that an isotropic emission occurs for the magic angle $\alpha_{iso}=$ atan$(\sqrt{2})=\SI{54.7}{\degree}$ corresponding to $A=B$.

\subsubsection{Analysis of the polarized emission}
\label{sect:lowNA}
\begin{table*}[ht] 
\caption{\label{MDangles}Angles $\alpha$ with respect to the c-axis of magnetic dipoles deduced from spectral analysis of single NaYF$_4$:Eu$^{3+}$ nanorods. Two models are considered: the paraxial approximation (parax., Eq. \ref{eq:paraxial}) and the full model (full, Eq. \ref{eq:polMD}) with NA=0.6. The uncertainties with a 95\% confidence interval are estimated from the fits.  The mean and the standard deviation are also indicated in the last rows.}
\begin{ruledtabular}
\begin{tabular}{l  | c c | c c | c c}
	& \multicolumn{2}{c|}{MD1 angle} &  \multicolumn{2}{c|}{MD2 angle} &  \multicolumn{2}{c}{MD3 angle}\\
Rod & parax. & full & parax. & full & parax. & full\\
\hline\hline
1 & 65.7$\pm0.1^\circ$ & 67.6$\pm0.3^\circ$ & 70.0$\pm0.2^\circ$ & 73.3$\pm0.3^\circ$ & 37.3$\pm0.3^\circ$ & 35.9$\pm0.4^\circ$ \\
2 & 66.0$\pm0.1^\circ$ & 68.1$\pm0.3^\circ$ & 69.0$\pm0.3^\circ$ & 73.3$\pm0.5^\circ$ & 37.2$\pm0.3^\circ$ &  35.7$\pm0.3^\circ$ \\
3 & 66.6$\pm0.1^\circ$ & 69.4$\pm0.7^\circ$ & 69.7$\pm0.2^\circ$ & 73.5$\pm0.2^\circ$ &38.0$\pm0.2^\circ$ & 36.3$\pm0.6^\circ$\\
4 & 65.3$\pm0.2^\circ$ & 67.4$\pm0.4^\circ$ & 67.7$\pm0.2^\circ$ & 70.9$\pm0.3^\circ$ & 38.4$\pm0.9^\circ$ & 36.9$\pm0.9^\circ$ \\
5 & 65.0$\pm0.1^\circ$ & 66.8$\pm0.2^\circ$ & 68.3$\pm0.1^\circ$ &  71.3$\pm0.1^\circ$ & 38.3$\pm0.2^\circ$ & 36.7$\pm0.3^\circ$  \\
\hline
mean & 65.7$^\circ$ & 67.9$^\circ$ & 68.9$^\circ$ & 72.5$^\circ$ & 37.8$^\circ$ & 36.3$^\circ$\\
std & 0.6$^\circ$ & 1.0$^\circ$ & 1.0$^\circ$ & 1.3$^\circ$ & 0.6$^\circ$ & 0.5$^\circ$\\
 \end{tabular}
\end{ruledtabular}
\end{table*}

\begin{figure*}
	\includegraphics[width=17cm]{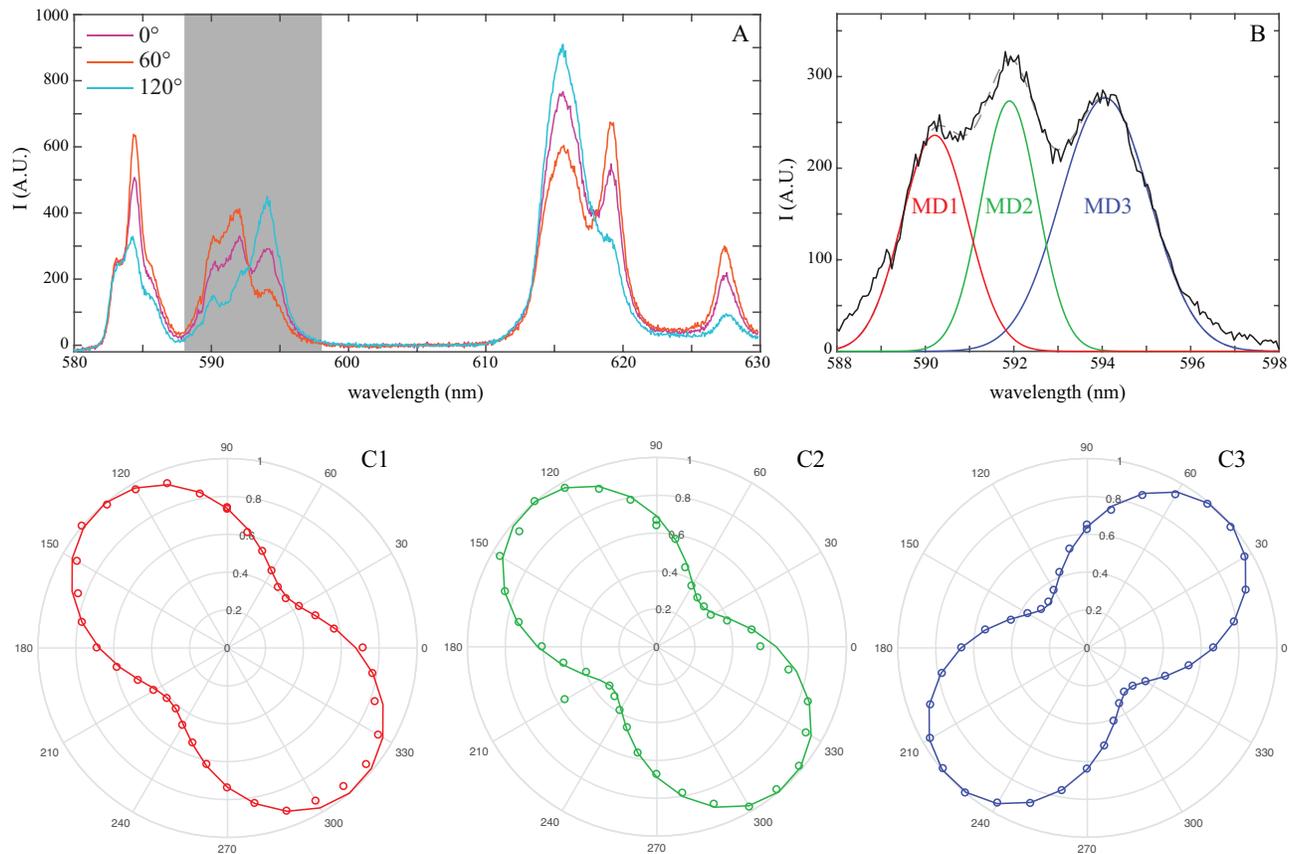}
	\caption{(a) Luminescence specra of single NaYF$_4$:Eu$^{3+}$ nanorod for different analyzer angles.  The $^5$D$_0\rightarrow ^7$F$_1$ magnetic transitions are highlighted in gray and zoomed in (b) for an analyzer angle of 0$^\circ$. Spectral unmixing  is performed by the fitting experimental data with a sum of three Gaussians. Three magnetic dipoles can be isolated, they are noted MD1 (in red, gaussian width 1.2 nm), MD2 (in green, gaussian width 0.8 nm) and MD3 (in blue, gaussian width 1.5 nm). The normalized polarization diagrams of MD1, MD2 and MD3 are shown with dots in the polar plots (c1), (c2) and (c3), respectively. For each dipole, a normalization is performed on the maximum emission of single nanorod. The lines correspond to fits using the full model.
		\label{fig:RodSpectra}}
\end{figure*}

The emission of single nanorod are acquired with the home made epi-confocal microscope described in section \ref{sect:ED-MD} and equipped with the low NA objective to stay within the paraxial approximation discussed above.  The emission spectra are recorded by varying the analyzer angle between 0$^\circ$  and 360$^\circ$ with a step of 10$^\circ$ and are shown in Fig. \ref{fig:RodSpectra}(a).

Let us focus on the $^5$D$_0\rightarrow ^7$F$_1$ magnetic transitions occurring between 588 nm and 598 nm (gray region). In this wavelength range, three main peaks are clearly visible. They are reproduced in Fig. \ref{fig:RodSpectra}(b).
For unmixing the proportion of each transition, the spectra are fitted with a sum of three Gaussian functions, represented in red, green and blue. We observe relatively narrow lines ($\sim 1$ nm) thanks to the single crystalline NaYF$_4$ host matrix, so that the three MD transitions can be well separated at room temperature and a full understanding of the polarization diagram can be proposed.  Based on the fitting amplitudes obtained for every analyzer angle, which are proportional to the total integral for a Gaussian, the normalized polarization diagrams of each of the transitions can be plotted. They are indicated with open circles in Fig. \ref{fig:RodSpectra}(c). Note that the luminescence emitted by magnetic transitions 1 and 2 (noted MD1 and MD2) have a preferred polarization aligned with the c-axis, whereas the luminescence from magnetic transition 3 (MD3) is  emitted perpendicularly to this axis. As already shown by Kim et al. \cite{Kim-Gacoin:17}, the orientation angle of the nanorod can now be determined and the angle of each magnetic dipole (from the c-axis) can be extracted by using the paraxial approximation (see Eq. \ref{eq:paraxial}).

We determine all these parameters (dipole angle and nanorod angle) by fitting the experimental points with the full theoretical model detailed in appendix \ref{sect:Green}, see Eq. \ref{eq:polMD}.  For rod \#4, whose recorded spectra are displayed in Fig. \ref{fig:RodSpectra}(a), the angles between each dipoles and the c-axis obtained in the paraxial approximation are equal to 65.3$\pm$0.2$^\circ$ for MD1 (in red), 67.7$\pm$0.2$^\circ$ for MD2 (in green) and 38.4$\pm$0.9$^\circ$ for MD3 (in blue). With the full model these angles become: 67.4$\pm$0.4$^\circ$ for dipole 1, 70.9$\pm$0.3$^\circ$ for MD2 and 36.9$\pm$0.9$^\circ$ for dipole MD3.
The difference between the angles deduced from the paraxial approximation and the full model remains small (less than 3$^\circ$). 

This procedure has been applied to several individual nanorods. The results are summarized in Table \ref{MDangles}. For all nanorods,  the small deviation between the values extracted from the paraxial approximation and the full model confirms that the paraxial theory is valid for a numerical aperture lower than 0.7 \cite{Sheppard-Matthews:87}. The small standard deviation (less than 1.5$^\circ$) indicates a minor dispersion of the angles. We conclude that the deduced dipole angles do not dependent on the chosen nanorod because these angles reflect the dipolar geometry resulting from the crystal field splitting of sublevels.

\section{Fourier plane leakage radiation microscopy}
\label{sect:Fourier}
The characterization of a single nanorod dipolar emission may also be performed with an other method called Fourier plane microscopy \cite{Drezet2008,GrandidierJMicrosc:2010,Taminiau-Zia:2012}. 
Unlike spectroscopic measurement based on a low NA detection, this method uses a high numerical aperture oil immersion objective (NA 1.49, Nikon) and  a camera  adequately positioned in a conjugate Fourier plane of the microscope.
The same continuous laser of wavelength 392 nm was used and a power of $\SI{287.5}{\micro \watt}$ (at the focal point) is used to excite Europium ions. The luminescence  is separated from the excitation laser with the same dichroic mirror, passed through a rotating analyzer and is detected with a scientific CMOS camera (Andor Zyla) placed in the Fourier plane. To achieve a good signal to noise ratio, we use a binning of $4\times 4$ and an integration times of 15 s. Two distinct band pass filters are used: from 586.8 to 593.6 nm (noted BP1) selecting equivalently MD1 and MD2 (proportions $p_1 \approx  p_2 \approx 50\%$), and  from 591 nm to 598.6 nm (noted BP2) selecting predominantly MD3 ($p_3\approx 88\%$) compared to MD2 ($p_2\approx 12\%$).

\begin{figure*}
	\includegraphics[width=18cm]{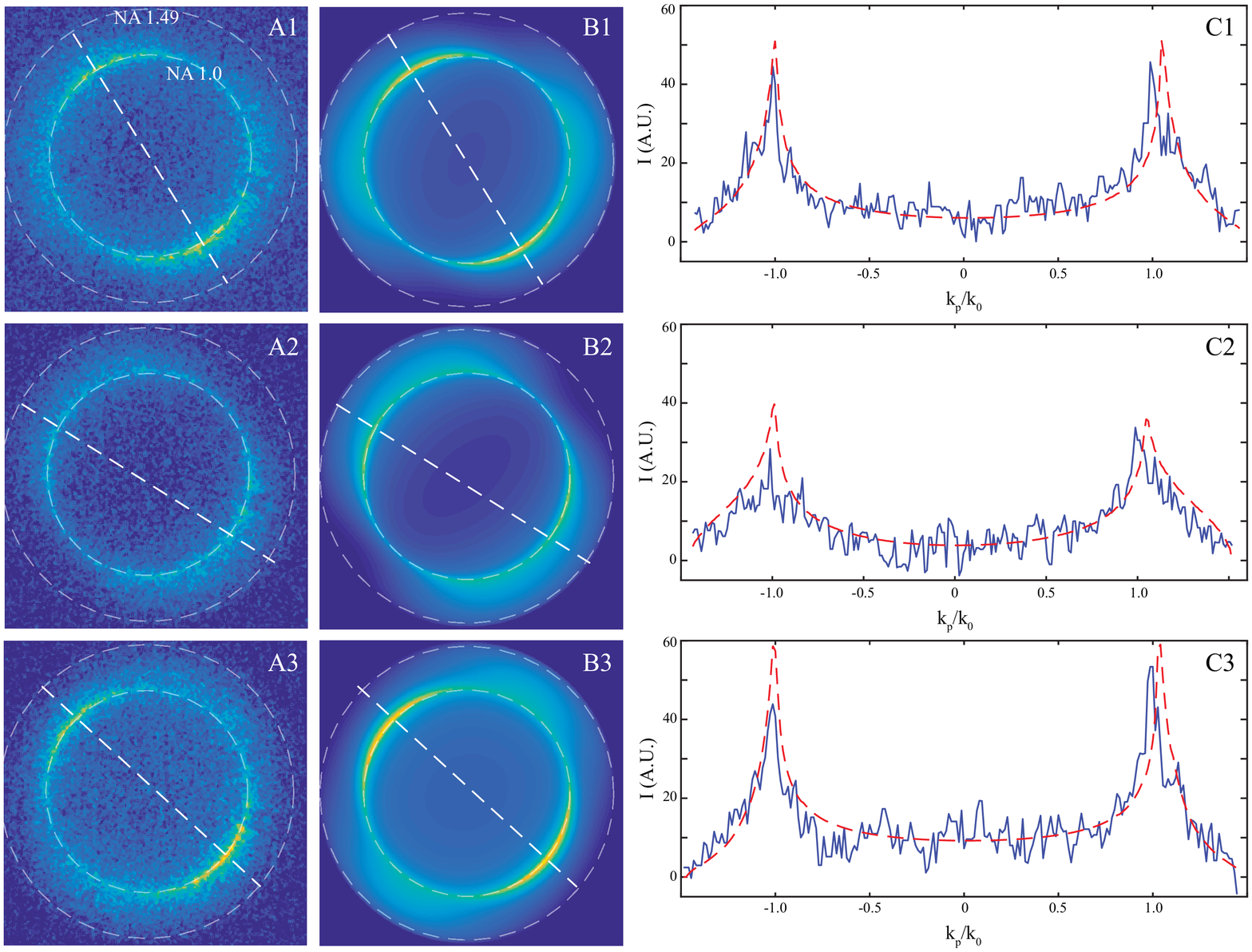}
	\caption{Fourier imaging of single NaYF$_4$:Eu$^{3+}$ nanorod for 586.8-593.6 nm wavelength range. Experimental Fourier images acquired with analyzer angle of 0$^\circ$, 60$^\circ$ and 120$^\circ$ are shown respectively  in a1, a2 and a3. The white circle with larger diameter indicates the maximum collection angle corresponding to the numerical aperture of the objective (NA=1.49). We also represent the circle indicating the critical angle for an air-quartz interface (corresponding to NA=1.0). These experimental images are compared to the theoretical ones shown in b1, b2 and b3 for an analyzer angle of respectively 0$^\circ$, 60$^\circ$ and 120$^\circ$. For each analyzer angle, we have represented in c1, c2 and c3, the experimental (in solid blue line) and theoretical (in dashed red line) intensity profiles corresponding to the white dashed lines in a) and b).
		\label{fig:FourierMD1}}
\end{figure*}

Typical emission patterns from a single nanorod with BP1 are shown in Fig. \ref{fig:FourierMD1}-(a1), (a2) and (a3) for three orientations of the analyzer. We observe two distinct lobes of higher intensity  at an angle near the critical angle $\theta_c$, with $\theta_c=43.3^\circ$ for an air-quartz interface. These two lobes slightly rotate with the analyzer angle (see video 1) and they are almost in the same direction as the rod axis, which is in agreement with the polarization emission plots discussed in Fig. \ref{fig:RodSpectra}-(c1) and (c2). A normalization of the intensity is performed on the maximal emission measured for all analyzer angles.

These experimental Fourier images can be fitted with the full theoretical model detailed in appendix \ref{sect:Green} (see Eq. \ref{eq:FourierMD}). The calculated Fourier plane image are shown in \ref{fig:FourierMD1}-(b1), (b2) and (b3), respectively. The agreement  between the experimental and simulated Fourier images is very good both on the overall distribution of the intensity and the angular position of the lobes as confirmed by the intensity profiles in Fig. \ref{fig:FourierMD1}-(c1), (c2) and (c3). From the simulated Fourier planes; we are in position to determine the angles between the magnetic dipoles MD1, MD2 and the c-axis as well as the proportion $p_1$. For rod \#4 for instance, we find angles of 67.4$^\circ$  and  70.8$^\circ$ for MD1 and MD2 respectively and a proportion of 50\%. These angles are almost identical to those obtained with the spectroscopic analysis (see previous section). 

\begin{table}[h!] 
	\caption{\label{MDFangles}Angles of magnetic dipoles deduced from Fourier images of single NaYF$_4$:Eu$^{3+}$ nanorods. Two band pass filters are considered: BP1 (586.8-593.6 nm) and BP2 (591.0-598.6 nm). Due to their large spectral widths, the emission of two dipoles are simultaneously collected. The proportion of MD1 (noted $p_1$), the proportion of MD3 (noted $p_3$) and the distance $z_0$ to the substrate are also estimated. The mean and the standard deviation are  indicated in the last rows.}
	\begin{ruledtabular}
		\begin{tabular}{l  | c c c | c c c |c | }
			& \multicolumn{3}{c|}{BP1} &  \multicolumn{3}{c|}{BP2} \\
			Rod & MD1 & MD2 & $p_1$  & MD2 & MD3 & $p_3$ & $z_0$ (nm)\\
			\hline\hline
			1 & 68.2$^\circ$ & 73.6$^\circ$ & 0.49  & 73.6$^\circ$ & 36.2$^\circ$ & 0.87 & 65.6 \\
			2 & 67.9$^\circ$ & 70.6$^\circ$ & 0.50 & 72.9$^\circ$ & 35.1$^\circ$ & 0.89 & 50.1 \\
			3 & 69.6$^\circ$ & 73.6$^\circ$ & 0.50 & 73.4$^\circ$ & 36.6$^\circ$ & 0.88 & 61.4 \\
			4 & 67.4$^\circ$  & 70.8$^\circ$ & 0.50  &69.9$^\circ$& 35.3$^\circ$ & 0.90 & 57.8\\
			5 & 66.5$^\circ$ & 71.2$^\circ$ & 0.50 & 71.2$^\circ$ & 36.9$^\circ$ & 0.88 & 63.2 \\
			\hline
			mean & 67.9$^\circ$ & 72.0$^\circ$ & 0.50 & 72.2$^\circ$ & 36.0$^\circ$ & 0.88 & 59.6\\
			std & 1.1$^\circ$ & 1.5$^\circ$ & 0.01 & 1.6$^\circ$ & 0.8$^\circ$ & 0.01 & 6.0
		\end{tabular}
	\end{ruledtabular}
\end{table}

\begin{figure}
	\includegraphics[width=9cm]{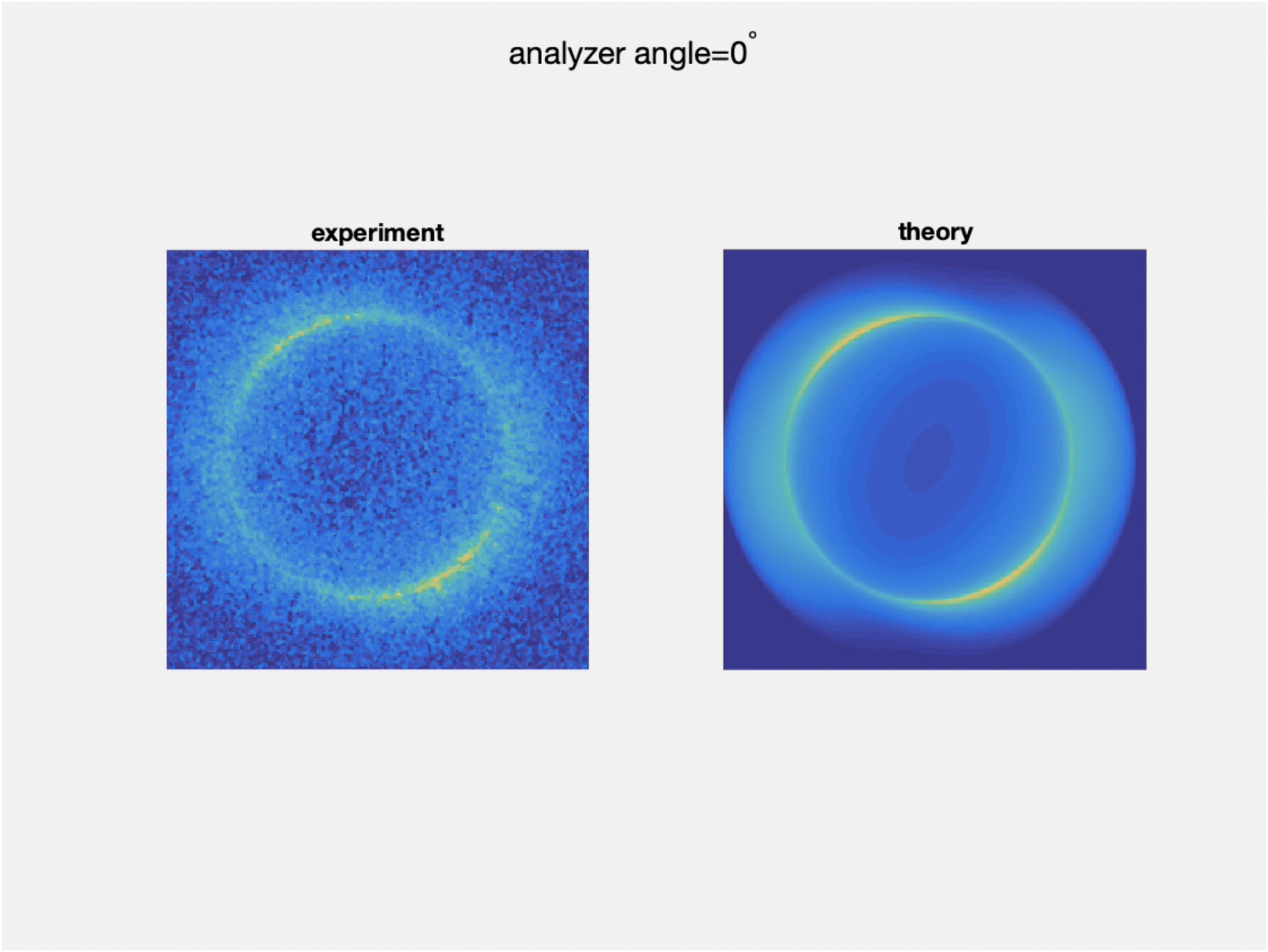}
	\caption{Video1. Fourier imaging of single NaYF$_4$:Eu$^{3+}$ nanorod for 586.8-593.6 nm wavelength range. Left) Experimental and right) calculated Fourier plane images as a function of analyzer angle from 0$^\circ$ to 180$^\circ$ by step of 30$^\circ$.
		\label{movie1}}
\end{figure}

\begin{figure*}[h!]
	\includegraphics[width=18cm]{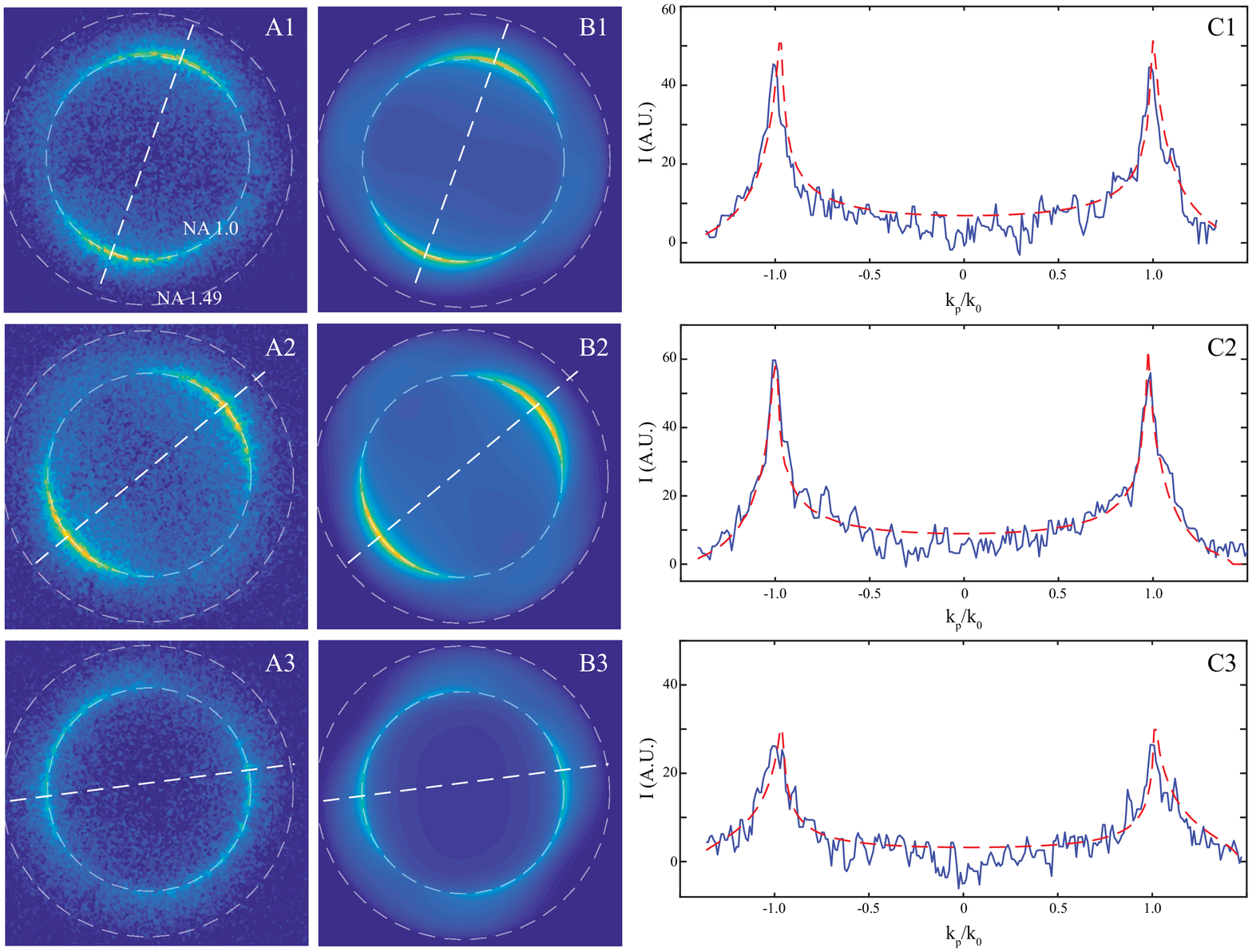}
	\caption{Fourier imaging of single NaYF$_4$:Eu$^{3+}$ nanorod for 591.0-598.6 nm wavelength range. Experimental Fourier images acquired with analyzer angle of 0$^\circ$, 60$^\circ$ and 120$^\circ$ are shown respectively  in a1, a2 and a3. The white circle with larger diameter indicates the maximum collection angle corresponding to the numerical aperture of the objective (NA=1.49). We also represent the circle indicating the critical angle for an air-quartz interface (corresponding to NA=1.0). These experimental images are compared to the theoretical ones shown in b1, b2 and b3 for an analyzer angle of respectively 0$^\circ$, 60$^\circ$ and 120$^\circ$. For each analyzer angle, we have represented in c1, c2 and c3, the experimental (in solid blue line) and theoretical (in dashed red line) intensity profiles corresponding to the white dashed lines in a) and b. 
		\label{fig:FourierMD2}}
\end{figure*}

We have applied the same procedure for analyzing the Fourier images acquired with the second band pass filter (BP2) shown in Fig. \ref{fig:FourierMD2}-(a1), (a2) and (a3). The two lobes rotate with the analyzer angle (see video 2) and are now perpendicular to the c-axis, as expected from Fig. \ref{fig:RodSpectra}(c3). By fitting these experimental images with the full theoretical model, we find angles of 69.9$^\circ$  and  35.3$^\circ$ for MD2 and MD3 respectively and a MD3 proportion of 90\%. Again, these angles are close to those obtained in the previous section. 

The results obtained for several individual nanorods  are summarized in Table \ref{MDFangles}. The angles are very similar to those obtained with full model used in conjunction of the spectral analysis, confirming that this method is robust for determining dipolar emission angles. 
The variability of the angle measurements is small and within the experimental error, estimated to 2$^\circ$ \cite{lieb04}, confirming that it does not depend on the chosen rod but reveals the local site symmetry of Europiums ions.

We emphasize that we fully explain the recorded Fourier images considering a fully magnetic transition but taking into account the contribution of the two MD transitions within the detected spectral range. Differently, several works considered randomly oriented emitters in solution or thin films and observe that the MD transition could present an electric dipole character \cite{Freed-Weissman:1941,Kunz-Lukosz:80,Taminiau-Zia:2012}, as discussed in the introduction. The ED contribution could originate from {\it e.g} the perturbation of the selection rules depending on the host matrix or from the contribution  
of the neighbor $^5$D$_1\rightarrow ^7$F$_3$ ED line \cite{Chang-Gruber:64}. Further investigations are necessary to conclude. For instance, low temperature measurement would help separating the emission lines. 

Furthermore, Fourier imaging performed with a high NA objective collects light emitted towards angles larger than the critical angle. Under this condition, the distance between the nanorod and the substrate (noted $z_0$) is critical and should be considered as  an additional fitting parameter in the model. As indicated in Table \ref{MDFangles}, we find an effective height of 59.6 nm which is in agreement with the dimension of the nanorod. This effective height is due to the finite diameter of the rod and consequently an integrated signal pickup over a finite rod volume. More precisely, the dipolar emission above the critical angle originates from the evanescent coupling into the substrate $e^{-\kappa_1 z_0} \;,  \kappa_1=\sqrt{\epsilon_3 \sin^2 \theta-\epsilon_1}$ for $\theta > \theta_c$ (see Eq. \ref{eq:GreenMag}). Therefore, high NA Fourier microscopy gives access to the effective height of the rod onto the substrate. It represents the ponderation by the evanescent coupling over the rod height and can be used to extract information on the distance between the rod and the substrate. 

\begin{figure}
	\includegraphics[width=9cm]{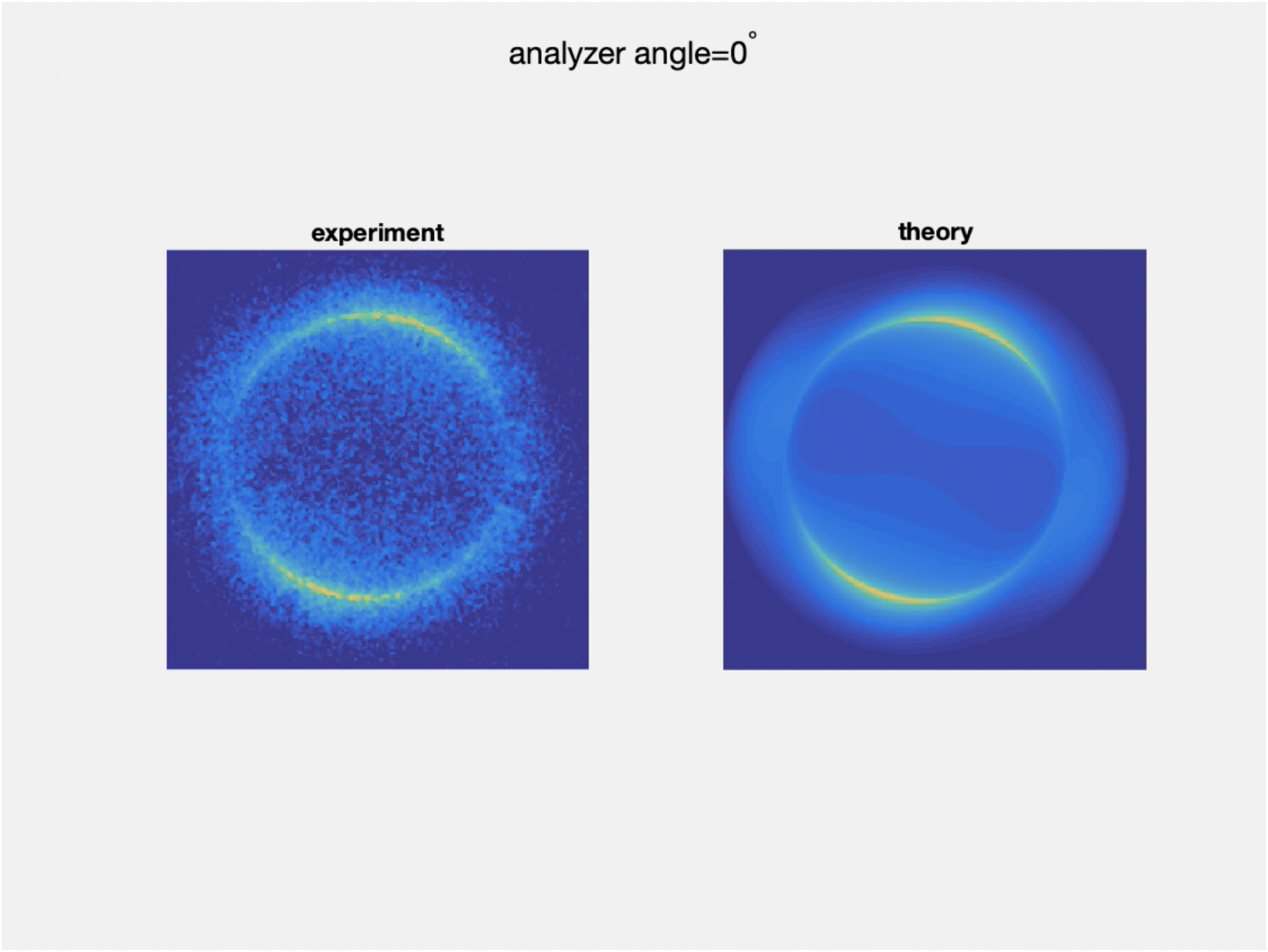}
	\caption{Video 2. Fourier imaging of single NaYF$_4$:Eu$^{3+}$ nanorod for 591.0-598.6 nm wavelength range. Left) Experimental and right) calculated Fourier plane images as a function of analyzer angle from 0$^\circ$ to 180$^\circ$ by step of 30$^\circ$.
		\label{movie2}}
\end{figure}

\section{Conclusion}
In summary, we fully characterized the magnetic dipole moment associated to the $^5$D$_0\rightarrow ^7$F$_1$ transition in individual  NaYF$_4$:Eu$^{3+}$ nanocrystals by two different approaches.  We determine the orientation of the three magnetic dipoles for the Stark sublevels originating from the crystal field degeneracy breaking. The orientation of the dipole moment is an intrinsic property of the synthesized nanorods and can be used for measuring their absolute orientation \cite{RodriguezSevilla:16,Kim-Gacoin:20}. In addition, since the optical transition properties are fully characterized and presents low variability from one rod to another, such nanocrystals could be used for mapping of the magnetic field and LDOS near nanophotonics structures, notably metamaterials and nano-antennas. In particular, fluorescence lifetime of rare-earth doped nanoparticles probes the electric or magnetic LDOS in complex environnement \cite{Karaveli-Zia:2011,Aigouy:14,Rabouw-Norris:16,Li-Zia:18}. Random orientation of the dipolar emitters is generally considered so that lifetime measures LDOS without information on the mode polarization. We expect additional information on the mode polarization considering the fixed dipole moment we characterized in NaYF$_4$:Eu$^{3+}$ single crystalline nanorods. This will be investigated in a future work. Using a high NA objective to pick up frustrated evanescent wave contribution of the dipolar emission into the substrate, we also determine an effective height of the rod dipole moment. The sensitivity of the Fourier imaging with respect to the altitude of the rod is also of interest to calibrate near-field optical measurement. It is worth mentioning that using a single crystalline NaYF$_4$ host matrix is a key parameter in this work since it leads to very narrow and well separated emission lines at room temperature clarifying the role of the Stark sublevels in the luminescence signal.  Another significant advantage is the ability to synthesize this compound as nanorods with very good control of size and shape, thus allowing to easily characterize and control the spatial orientation of the crystal and the crystalline axis, for full engineering of their spectroscopic properties at the nanoscale level.

\begin{acknowledgments}
We gratefully acknowledge Fr\'ed\'eric Herbst from the technological platform ARCEN Carnot for SEM imaging of single nanorods. The platform ARCEN Carnot is financed by the R\'egion de Bourgogne Franche-Comt\'e and the D\'el\'egation R\'egionale \`a la Recherche et \`a la Technologie (DRRT). 
This work is supported by the French Investissements d'Avenir program EUR-EIPHI (17-EURE-0002) and  French National Research Agency (ANR) project SpecTra (ANR-16-CE24-0014).
\end{acknowledgments}
\appendix
\section{Synthesis of  single crystalline NaYF$_4$:Eu$^{3+}$nanorods}
\label{sect:Synthesis}
45 mmol (1.8 g) of NaOH in 6 mL of water were mixed with 15 mL of ethanol (EtOH) and 30 mL of oleic acid (OA) under stirring. To the resulting mixture were selectively added 0.95 mmol (288 mg) of Y(Cl)$_3$,6H$_2$O, 0.05 mmol (18 mg) of Eu(Cl)$_3$,6H$_2$O and 10.2 mmol (377mg) of NH$_4$F dissolved in 4 mL of water. The solution was then transferred into a 75 mL autoclave and heated at $\SI{200}{\celsius}$  for 24 h under stirring. After cooling down to ambient temperature, the resulting nanoparticles were precipitated by addition of 50 mL of ethanol, collected by centrifugation, and washed three times with water and ethanol. They were finally dried under vaccum and recovered as a white powder. 
For the optical experiments, a functionalization by ligand exchange is needed to ensure the good dispersion of the particles in water. About 20 mg of the NaYF$_4@$oleic acid NPs are dispersed in 2 ml of an aqueous 0.2 M sodium citrate solution. After sonication and centrifugation, the process of ligand exchange is repeated two more times. Finally, particles are washed three times with water and ethanol, and dispersed with sonication in pure water. Sample for optical characterizations are then prepared by spin-coating a droplet of this rod solution (concentration 0.014\% vol.) on a quartz substrate at 2000 rpm during 60s.

\section{Arbitrary rod orientation}
\label{sect:caxisBeta}
For arbitrary orientation of the rod on the substrate, we define  the angle $\beta$ with respect to the normal to the substrate ($\beta=\pi/2$ for a rod c-axis along the substrate). The dipole moment associated to a transition line presents a fixed angle $\alpha$ with respect to the c-axis, and is expressed ${\bf p}=p_0{\cal {\bf R}}_y(\beta) (\sin \alpha \cos \psi,\sin \alpha \sin \psi, \cos \alpha)$, where ${\cal {\bf R}}_y$ refers to the rotation around the y-axis. The magnetic field scattered at ${\bf r}$ by a dipole located at ${\bf r_0}$ can be expressed thanks to the magnetic Green's tensor ${\bf  Q}$
\begin{eqnarray} 
{\bf H}({\bf r})=\mu_0\omega_0^2 {\bf  Q}({\bf r},{\bf r_0},\omega_0) \cdot {\bf p}
\end{eqnarray}
and the incoherent emission of all the emitting dipole (for $\psi \in [0; 2\pi[$) can be expressed
\begin{widetext}
\begin{eqnarray} 
&&{\bf I}({\bf r})=\frac{1}{2\pi n_3}\int_0^{2\pi}\left \vert {\bf H}({\bf r})\right \vert ^2 d\psi
\propto  \left( \left \vert Q_{xx}\right \vert ^2+\left \vert Q_{yx}\right \vert ^2+\left \vert Q_{zx}\right \vert ^2\right) \left( \frac{1}{2}\cos^2\beta\sin^2\alpha+\sin^2\beta\cos^2\alpha \right)\\
\nonumber
&&+ \left( \left \vert Q_{xy}\right \vert ^2+\left \vert Q_{yy}\right \vert ^2+\left \vert Q_{zy}\right \vert ^2\right) \frac{1}{2}\sin^2\alpha 
+ \left( \left \vert Q_{xz}\right \vert ^2+\left \vert Q_{yz}\right \vert ^2+\left \vert Q_{zz}\right \vert ^2\right)  \left( \frac{1}{2}\sin^2\beta\sin^2\alpha s+\cos^2\beta\cos^2\alpha \right) \\
\nonumber
&&+2Re\left( Q_{xx}G_{xy}+Q_{yx}G_{xz}+Q_{zx}Q_{zz}\right)\times  \left(-\frac{1}{2}\cos\beta\sin\beta \sin^2\alpha +\sin\beta\cos \beta \cos^2\alpha \right)
\end{eqnarray}
that is equivalent to the incoherent emission of three orthogonal dipoles 
\begin{eqnarray}
\nonumber
{\bf p}_1&=&
\begin{pmatrix}
\sin\beta \cos\alpha \\0 \\ \cos \beta \cos\alpha 
\end{pmatrix} \,,
{\bf p}_2=\frac{1}{\sqrt{2}}
\begin{pmatrix}
0 \\ \sin \alpha \\0 
\end{pmatrix} \,,
{\bf p}_3=\frac{1}{\sqrt{2}}
\begin{pmatrix}
-\cos\beta \sin\alpha \\0 \\ \sin \beta \sin\alpha 
\end{pmatrix} \,.
\end{eqnarray}
\end{widetext}
Analogous expressions hold for the electric field emitted by an ED. 
\section{Far-field ED and MD scattering}
\label{sect:Green}
\begin{figure}
	\includegraphics[width=8cm]{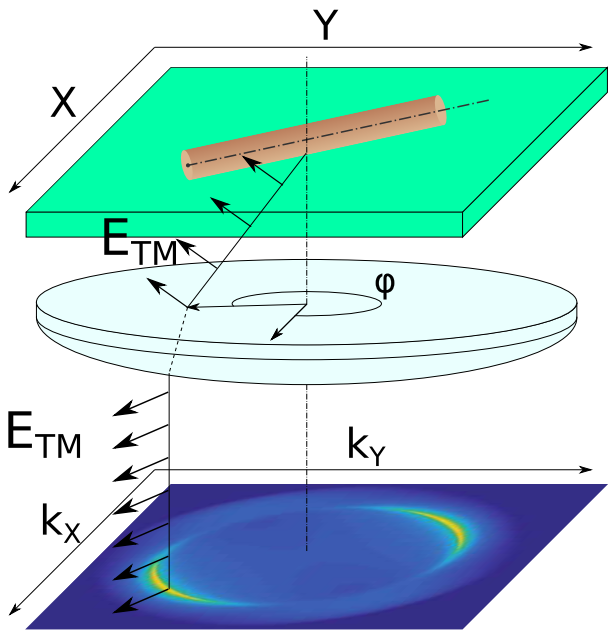}
	\caption{Notation for the scattered dipolar field.
		\label{fig:RodDipoleConfocal}}
\end{figure}
\subsection{Electric dipole}
\subsubsection{Electric field Green's tensor} 
The electric field scattered at the position ${\bf r}$ from an ED located at ${\bf r}_0$ is 
\begin{eqnarray} 
{\bf E}({\bf r})=\frac{k_0^2}{\epsilon_0} {\bf  G}({\bf r},{\bf r_0},\omega_0) \cdot {\bf p}
\end{eqnarray}
where ${\bf  G}$ is the electric field Green's tensor. For $k_0 r \gg 1$, far-field appromixation holds and \cite{Novotny-Hecht:2006}
\begin{eqnarray}
\label{eq:GreenElec}
&{\bf G}({\bf r},{\bf r}_0) \underset{r \to \infty}{\sim}  \frac{n_3}{n_1} \frac{e^{ik_3 r}}{r}  \cos \theta  e^{iw_1 z_0} {\bf g}
\nonumber
\end{eqnarray}
with $k_1=n_1k_0\;, k_3=n_3k_0 \;,w_1=(k_1^2-k_3^2 \sin^2 \theta)^{1/2} \;, [Im (w_1) \ge 0]$ and 
\begin{eqnarray*}
g_{xx}&=&t_p \cos \theta \cos ^2\varphi-\frac{k_1}{w_1}t_s\sin^2 \varphi \\
g_{xy}&=&t_p\cos\theta \cos\varphi\sin \varphi +\frac{k_1}{w_1}t_s \cos \varphi \sin \varphi \\
g_{xz}&=&\frac{k_3}{w_1}t_p \cos \theta\sin \theta  \cos \varphi \\
g_{yx}&=&g_{xy}\\
g_{yy}&=&t_p \cos \theta \sin ^2\varphi-\frac{k_1}{w_1}t_s\cos^2 \varphi \\
g_{yz}&=&\frac{k_3}{w_1}t_p \cos \theta\sin \theta  \sin \varphi \\
g_{xz}&=&-t_p \sin \theta  \cos \varphi \\
g_{yz}&=&-t_p \sin \theta  \sin \varphi \\
g_{yz}&=&-\frac{k_3}{w_1}t_p \sin^2 \theta 
\end{eqnarray*}
$t_s$ ($t_p$) refer to the fresnel air/glass transmission coefficient for TE (TM) polarized light.
\subsubsection{Polarized emission}
The emission polarization is analysed by confocal microscopy. For numerical simulations, we first decompose the electric field in TE and TM polarization. 
\begin{eqnarray}
{\bf E}=E_{TM}{\bf u}_p+E_{TE} {\bf u}_s
\end{eqnarray}
 where we define the unitary vectors ($k_p^2=k_x^2+k_y^2$)
\begin{eqnarray}
{\bf u}_p=\frac{1}{k_3k_p}
\begin{pmatrix}
k_x w_3 \\
k_y w_3\\
-k_p^2
\end{pmatrix}
\;;
{\bf u}_s=\frac{1}{k_p}
\begin{pmatrix}
k_y \\
-k_x\\
0
\end{pmatrix}
\end{eqnarray}
After refraction on the reference sphere (see Fig. \ref{fig:RodDipoleConfocal}), the components of the electric field are 
\begin{eqnarray}
E_{fx}&=&-E_{TM}\cos \varphi+E_{TE}\sin\varphi\\
E_{fy}&=&-E_{TM}\sin \varphi-E_{TM}\cos\varphi
\end{eqnarray}
Finally, if we note $\Phi_A$ the angle of the analyzer with respect to the x-axis, the polarized intensity in the Fourier plane is expressed as
\begin{eqnarray}
I(\theta,\varphi)=\frac{n_3}{\vert cos \theta\vert} \lvert \cos(\Phi_A)E_{fx}+\sin(\Phi_A)E_{fy}\rvert^2
\end{eqnarray}
where the apodization factor $1/\vert \cos \theta \vert$ ensures the energy conservation. 
In addition, the polarization diagram is obtained after integration over all the numerical aperture ($\theta_{NA}=asin(NA/n_3)$).
\begin{eqnarray}
I_{pol}(\Phi_A)=\int_\pi^{\pi+\theta_{NA}} d\theta\int_0^{2\pi} d\varphi I(\theta,\varphi) \vert \cos\theta\sin\theta \vert  
\end{eqnarray}
For detector located on the confocal optical axis ($\theta=\pi, \varphi=0$), $t_p=t_s$ and the  incoherent emission of the three orthogonal dipoles
\begin{eqnarray}
\nonumber
{\bf p}_1&=&
\begin{pmatrix}
\cos\alpha \\0 \\ 0
\end{pmatrix} \,,
{\bf p}_2=\frac{1}{\sqrt{2}}
\begin{pmatrix}
0 \\ \sin \alpha  \\0 
\end{pmatrix} \,,
{\bf p}_3=\frac{1}{\sqrt{2}}
\begin{pmatrix}
0 \\0 \\ \sin\alpha
\end{pmatrix} \,.
\end{eqnarray}
simplifies to  
\begin{eqnarray}
I_{pol}(\Phi_A)&\propto&A\sin^2\Phi_A +B\cos^2\Phi_A \;, \text{with}\\
A&=&\frac{\sin^2 \alpha}{2} \\
B&=&\cos^2 \alpha 
\end{eqnarray}
\subsection{Magnetic dipole} 
For a magnetic dipole, we have the analogous expressions 
\begin{eqnarray}
&& {\bf H}({\bf r})=\mu_0\omega^2{\bf  Q}({\bf r},{\bf r_0},\omega_0) \cdot {\bf p} 
\end{eqnarray}
with the magnetic Green tensor
\begin{eqnarray}
\label{eq:GreenMag}
&&{\bf Q}({\bf r},{\bf r}_0) \underset{r \to \infty}{\sim}  \epsilon_3 \frac{e^{ik_3 r}}{r}  \cos \theta e^{iw_1 z_0}  {\bf q}
\end{eqnarray}
\begin{eqnarray*}
q_{xx}&=&t_s \cos \theta \cos ^2\varphi-\frac{k_1}{w_1}t_p\sin^2 \varphi \\
q_{xy}&=&t_s\cos\theta \cos\varphi\sin \varphi +\frac{k_1}{w_1}t_p \cos \varphi \sin \varphi \\
q_{xz}&=&\frac{k_3}{w_1}t_s \cos \theta\sin \theta  \cos \varphi \\
q_{yx}&=&q_{xy}\\
q_{yy}&=&t_s \cos \theta \sin ^2\varphi-\frac{k_1}{w_1}t_p\cos^2 \varphi \\
q_{yz}&=&\frac{k_3}{w_1}t_s \cos \theta\sin \theta  \sin \varphi \\
q_{xz}&=&-t_s \sin \theta  \cos \varphi \\
q_{yz}&=&-t_s \sin \theta  \sin \varphi \\
q_{yz}&=&-\frac{k_3}{w_1}t_s \sin^2 \theta 
\end{eqnarray*}
The TE/TM decomposition follows
\begin{eqnarray}
{\bf H}=H_{TM}{\bf v}_p+H_{TE} {\bf v}_s
\end{eqnarray}
 with the unitary vectors
\begin{eqnarray}
{\bf v}_s=\frac{1}{k_3k_p}
\begin{pmatrix}
k_x w_3 \\
k_y w_3\\
-k_p^2
\end{pmatrix}
\;;
{\bf v}_p=\frac{1}{k_p}
\begin{pmatrix}
-k_y \\
k_x\\
0
\end{pmatrix} \;.
\end{eqnarray}
After refraction on the reference sphere (see Fig. \ref{fig:RodDipoleConfocal}), the components of the magnetic field are 
\begin{eqnarray}
H_{fx}&=&-H_{TE}\cos \varphi-H_{TM}\sin\varphi\\
H_{fy}&=&-H_{TE}\sin \varphi+H_{TE}\cos\varphi
\end{eqnarray}
Finally, the polarized intensity in the Fourier plane is expressed as
\begin{eqnarray}
I(\theta,\varphi)=\frac{1}{n_3\vert \cos \theta\vert} \lvert -\sin(\Phi_A)H_{fx}+\cos(\Phi_A)H_{fy}\rvert^2 \hspace{0.75cm}
\label{eq:FourierMD}
\end{eqnarray}
The polarized emission recorded in the confocal microscope of numerical aperture NA is 
\begin{eqnarray}
I_{pol}(\Phi_A)=\int_\pi^{\pi+\theta_{NA}} d\theta\int_0^{2\pi} d\varphi I(\theta,\varphi) \vert \cos\theta\sin\theta \vert  \hspace{0.75cm}
\label{eq:polMD}
\end{eqnarray}
In the paraxial approximation, the incoherent emission of the three orthogonal dipole defined above simplifies to  
\begin{eqnarray}
I_{pol}(\Phi_A)&=&A\sin^2\Phi_A +B\cos ^2\Phi_A \;, \text{with}\\
A&=&\cos^2 \alpha \\
B&=&\frac{\sin^2 \alpha}{2}
\end{eqnarray}


\begin{thebibliography}{34}%
\makeatletter
\providecommand \@ifxundefined [1]{%
 \@ifx{#1\undefined}
}%
\providecommand \@ifnum [1]{%
 \ifnum #1\expandafter \@firstoftwo
 \else \expandafter \@secondoftwo
 \fi
}%
\providecommand \@ifx [1]{%
 \ifx #1\expandafter \@firstoftwo
 \else \expandafter \@secondoftwo
 \fi
}%
\providecommand \natexlab [1]{#1}%
\providecommand \enquote  [1]{``#1''}%
\providecommand \bibnamefont  [1]{#1}%
\providecommand \bibfnamefont [1]{#1}%
\providecommand \citenamefont [1]{#1}%
\providecommand \href@noop [0]{\@secondoftwo}%
\providecommand \href [0]{\begingroup \@sanitize@url \@href}%
\providecommand \@href[1]{\@@startlink{#1}\@@href}%
\providecommand \@@href[1]{\endgroup#1\@@endlink}%
\providecommand \@sanitize@url [0]{\catcode `\\12\catcode `\$12\catcode
  `\&12\catcode `\#12\catcode `\^12\catcode `\_12\catcode `\%12\relax}%
\providecommand \@@startlink[1]{}%
\providecommand \@@endlink[0]{}%
\providecommand \url  [0]{\begingroup\@sanitize@url \@url }%
\providecommand \@url [1]{\endgroup\@href {#1}{\urlprefix }}%
\providecommand \urlprefix  [0]{URL }%
\providecommand \Eprint [0]{\href }%
\providecommand \doibase [0]{https://doi.org/}%
\providecommand \selectlanguage [0]{\@gobble}%
\providecommand \bibinfo  [0]{\@secondoftwo}%
\providecommand \bibfield  [0]{\@secondoftwo}%
\providecommand \translation [1]{[#1]}%
\providecommand \BibitemOpen [0]{}%
\providecommand \bibitemStop [0]{}%
\providecommand \bibitemNoStop [0]{.\EOS\space}%
\providecommand \EOS [0]{\spacefactor3000\relax}%
\providecommand \BibitemShut  [1]{\csname bibitem#1\endcsname}%
\let\auto@bib@innerbib\@empty
\bibitem [{\citenamefont {Shalaev}(2007)}]{Shalaev:07}%
  \BibitemOpen
  \bibfield  {author} {\bibinfo {author} {\bibfnamefont {V.}~\bibnamefont
  {Shalaev}},\ }\bibfield  {title} {\bibinfo {title} {Optical negative-index
  metamaterials},\ }\href@noop {} {\bibfield  {journal} {\bibinfo  {journal}
  {Nature Photonics}\ }\textbf {\bibinfo {volume} {1}},\ \bibinfo {pages} {41}
  (\bibinfo {year} {2007})}\BibitemShut {NoStop}%
\bibitem [{\citenamefont {Bidault}\ \emph {et~al.}(2019)\citenamefont
  {Bidault}, \citenamefont {Mivelle},\ and\ \citenamefont
  {Bonod}}]{Bidault-Mivelle-Bonod:19}%
  \BibitemOpen
  \bibfield  {author} {\bibinfo {author} {\bibfnamefont {S.}~\bibnamefont
  {Bidault}}, \bibinfo {author} {\bibfnamefont {M.}~\bibnamefont {Mivelle}},\
  and\ \bibinfo {author} {\bibfnamefont {N.}~\bibnamefont {Bonod}},\ }\bibfield
   {title} {\bibinfo {title} {Dielectric nanoantennas to manipulate solid-state
  light emission},\ }\href@noop {} {\bibfield  {journal} {\bibinfo  {journal}
  {Journal of Applied Physics}\ }\textbf {\bibinfo {volume} {126}} (\bibinfo
  {year} {2019})}\BibitemShut {NoStop}%
\bibitem [{\citenamefont {Devaux}\ \emph {et~al.}(2000)\citenamefont {Devaux},
  \citenamefont {Devaux}, \citenamefont {Bourillot}, \citenamefont {Weeber},
  \citenamefont {Lacroute}, \citenamefont {Goudonnet},\ and\ \citenamefont
  {Girard}}]{Devaux-Dereux-Bourillot-Weeber-Lacroute-Goudonnet-Girard:2000}%
  \BibitemOpen
  \bibfield  {author} {\bibinfo {author} {\bibfnamefont {E.}~\bibnamefont
  {Devaux}}, \bibinfo {author} {\bibfnamefont {E.}~\bibnamefont {Devaux}},
  \bibinfo {author} {\bibfnamefont {E.}~\bibnamefont {Bourillot}}, \bibinfo
  {author} {\bibfnamefont {J.~C.}\ \bibnamefont {Weeber}}, \bibinfo {author}
  {\bibfnamefont {Y.}~\bibnamefont {Lacroute}}, \bibinfo {author}
  {\bibfnamefont {J.~P.}\ \bibnamefont {Goudonnet}},\ and\ \bibinfo {author}
  {\bibfnamefont {C.}~\bibnamefont {Girard}},\ }\bibfield  {title} {\bibinfo
  {title} {Local detection of the optical magnetic field in the near zone of
  dielectric sample},\ }\href@noop {} {\bibfield  {journal} {\bibinfo
  {journal} {Physical Review B}\ }\textbf {\bibinfo {volume} {62}},\ \bibinfo
  {pages} {10504} (\bibinfo {year} {2000})}\BibitemShut {NoStop}%
\bibitem [{\citenamefont {Mary}\ \emph {et~al.}(2005)\citenamefont {Mary},
  \citenamefont {Dereux},\ and\ \citenamefont
  {Ferrell}}]{Mary-Dereux-Ferrell:2005}%
  \BibitemOpen
  \bibfield  {author} {\bibinfo {author} {\bibfnamefont {A.}~\bibnamefont
  {Mary}}, \bibinfo {author} {\bibfnamefont {A.}~\bibnamefont {Dereux}},\ and\
  \bibinfo {author} {\bibfnamefont {T.}~\bibnamefont {Ferrell}},\ }\bibfield
  {title} {\bibinfo {title} {Localized surface plasmons on a torus in the
  non--retarded approximation},\ }\href@noop {} {\bibfield  {journal} {\bibinfo
   {journal} {Physical Review B}\ }\textbf {\bibinfo {volume} {72}},\ \bibinfo
  {pages} {155426} (\bibinfo {year} {2005})}\BibitemShut {NoStop}%
\bibitem [{\citenamefont {Shafiei}\ \emph {et~al.}(2013)\citenamefont
  {Shafiei}, \citenamefont {Monticone}, \citenamefont {Le}, \citenamefont
  {Liu}, \citenamefont {T.Hartsfield}, \citenamefont {Alu},\ and\ \citenamefont
  {Li}}]{Shafei:13}%
  \BibitemOpen
  \bibfield  {author} {\bibinfo {author} {\bibfnamefont {F.}~\bibnamefont
  {Shafiei}}, \bibinfo {author} {\bibfnamefont {F.}~\bibnamefont {Monticone}},
  \bibinfo {author} {\bibfnamefont {K.}~\bibnamefont {Le}}, \bibinfo {author}
  {\bibfnamefont {X.-X.}\ \bibnamefont {Liu}}, \bibinfo {author} {\bibnamefont
  {T.Hartsfield}}, \bibinfo {author} {\bibfnamefont {A.}~\bibnamefont {Alu}},\
  and\ \bibinfo {author} {\bibfnamefont {X.}~\bibnamefont {Li}},\ }\bibfield
  {title} {\bibinfo {title} {A subwavelength plasmonic metamolecule exhibiting
  magnetic-based optical fano resonance},\ }\href@noop {} {\bibfield  {journal}
  {\bibinfo  {journal} {Nature Nanotechnology}\ }\textbf {\bibinfo {volume}
  {8}},\ \bibinfo {pages} {95} (\bibinfo {year} {2013})}\BibitemShut {NoStop}%
\bibitem [{\citenamefont {le~Feber}\ \emph {et~al.}(2013)\citenamefont
  {le~Feber}, , \citenamefont {Rotenberg}, \citenamefont {Beggs},\ and\
  \citenamefont {Kuipers}}]{leFeber-Kuipers:13}%
  \BibitemOpen
  \bibfield  {author} {\bibinfo {author} {\bibfnamefont {B.}~\bibnamefont
  {le~Feber}}, \bibinfo {author} {\bibfnamefont {N.}~\bibnamefont
  {Rotenberg}}, \bibinfo {author} {\bibfnamefont {D.}~\bibnamefont {Beggs}},\
  and\ \bibinfo {author} {\bibfnamefont {L.}~\bibnamefont {Kuipers}},\
  }\bibfield  {title} {\bibinfo {title} {Simultaneous measurements of
  nanoscale electric and magnetic optical fields},\ }\href@noop {} {\bibfield
  {journal} {\bibinfo  {journal} {Nature Photonics}\ }\textbf {\bibinfo
  {volume} {8}},\ \bibinfo {pages} {43} (\bibinfo {year} {2013})}\BibitemShut
  {NoStop}%
\bibitem [{\citenamefont {Taminiau}\ \emph {et~al.}(2012)\citenamefont
  {Taminiau}, \citenamefont {Karaveli}, \citenamefont {van Hulst},\ and\
  \citenamefont {Zia}}]{Taminiau-Zia:2012}%
  \BibitemOpen
  \bibfield  {author} {\bibinfo {author} {\bibfnamefont {T.}~\bibnamefont
  {Taminiau}}, \bibinfo {author} {\bibfnamefont {S.}~\bibnamefont {Karaveli}},
  \bibinfo {author} {\bibfnamefont {N.}~\bibnamefont {van Hulst}},\ and\
  \bibinfo {author} {\bibfnamefont {R.}~\bibnamefont {Zia}},\ }\bibfield
  {title} {\bibinfo {title} {Quantifying the magnetic nature of light
  emission},\ }\href@noop {} {\bibfield  {journal} {\bibinfo  {journal} {Nature
  Communications}\ }\textbf {\bibinfo {volume} {3}},\ \bibinfo {pages} {979}
  (\bibinfo {year} {2012})}\BibitemShut {NoStop}%
\bibitem [{\citenamefont {Blasse}\ and\ \citenamefont
  {Grabmaier}(1994)}]{Blasse-Grabmaier:94}%
  \BibitemOpen
  \bibfield  {author} {\bibinfo {author} {\bibfnamefont {G.}~\bibnamefont
  {Blasse}}\ and\ \bibinfo {author} {\bibfnamefont {B.}~\bibnamefont
  {Grabmaier}},\ }\href@noop {} {\emph {\bibinfo {title} {Luminescent
  Materials}}}\ (\bibinfo  {publisher} {Springer Verlag, Berlin},\ \bibinfo
  {year} {1994})\BibitemShut {NoStop}%
\bibitem [{\citenamefont {Vetrone}\ \emph {et~al.}(2010)\citenamefont
  {Vetrone}, \citenamefont {Naccache}, \citenamefont {Zamarr\'on},
  \citenamefont {Juarranz de~la Fuente}, \citenamefont {Sanz-Rodr\`iguez},
  \citenamefont {Martinez~Maestro}, \citenamefont {Mart\'in~Rodriguez},
  \citenamefont {Jaque}, \citenamefont {Garc\`ia~Sol\'e},\ and\ \citenamefont
  {Capobianco}}]{vetrone:10}%
  \BibitemOpen
  \bibfield  {author} {\bibinfo {author} {\bibfnamefont {F.}~\bibnamefont
  {Vetrone}}, \bibinfo {author} {\bibfnamefont {R.}~\bibnamefont {Naccache}},
  \bibinfo {author} {\bibfnamefont {A.}~\bibnamefont {Zamarr\'on}}, \bibinfo
  {author} {\bibfnamefont {A.}~\bibnamefont {Juarranz de~la Fuente}}, \bibinfo
  {author} {\bibfnamefont {F.}~\bibnamefont {Sanz-Rodr\`iguez}}, \bibinfo
  {author} {\bibfnamefont {L.}~\bibnamefont {Martinez~Maestro}}, \bibinfo
  {author} {\bibfnamefont {E.}~\bibnamefont {Mart\'in~Rodriguez}}, \bibinfo
  {author} {\bibfnamefont {D.}~\bibnamefont {Jaque}}, \bibinfo {author}
  {\bibfnamefont {J.}~\bibnamefont {Garc\`ia~Sol\'e}},\ and\ \bibinfo {author}
  {\bibfnamefont {J.~A.}\ \bibnamefont {Capobianco}},\ }\bibfield  {title}
  {\bibinfo {title} {Temperature sensing using fluorescent nanothermometers},\
  }\href {https://doi.org/10.1021/nn100244a} {\bibfield  {journal} {\bibinfo
  {journal} {ACS Nano}\ }\textbf {\bibinfo {volume} {4}},\ \bibinfo {pages}
  {3254} (\bibinfo {year} {2010})}\BibitemShut {NoStop}%
\bibitem [{\citenamefont {Saidi}\ \emph {et~al.}(2009)\citenamefont {Saidi},
  \citenamefont {Samson}, \citenamefont {Aigouy}, \citenamefont {Volz},
  \citenamefont {Low}, \citenamefont {Bergaud},\ and\ \citenamefont
  {Mortier}}]{Saidi:2009}%
  \BibitemOpen
  \bibfield  {author} {\bibinfo {author} {\bibfnamefont {E.}~\bibnamefont
  {Saidi}}, \bibinfo {author} {\bibfnamefont {B.}~\bibnamefont {Samson}},
  \bibinfo {author} {\bibfnamefont {L.}~\bibnamefont {Aigouy}}, \bibinfo
  {author} {\bibfnamefont {S.}~\bibnamefont {Volz}}, \bibinfo {author}
  {\bibfnamefont {P.}~\bibnamefont {Low}}, \bibinfo {author} {\bibfnamefont
  {C.}~\bibnamefont {Bergaud}},\ and\ \bibinfo {author} {\bibfnamefont
  {M.}~\bibnamefont {Mortier}},\ }\bibfield  {title} {\bibinfo {title}
  {Scanning thermal imaging by near-fieldfluorescence spectroscopy},\
  }\href@noop {} {\bibfield  {journal} {\bibinfo  {journal} {Nanotechnology}\
  }\textbf {\bibinfo {volume} {20}},\ \bibinfo {pages} {115703} (\bibinfo
  {year} {2009})}\BibitemShut {NoStop}%
\bibitem [{\citenamefont {Aigouy}\ \emph {et~al.}(2003)\citenamefont {Aigouy},
  \citenamefont {de~Wilde},\ and\ \citenamefont
  {Mortier}}]{Aigouy-deWilde-Mortier:2003}%
  \BibitemOpen
  \bibfield  {author} {\bibinfo {author} {\bibfnamefont {L.}~\bibnamefont
  {Aigouy}}, \bibinfo {author} {\bibfnamefont {Y.}~\bibnamefont {de~Wilde}},\
  and\ \bibinfo {author} {\bibfnamefont {M.}~\bibnamefont {Mortier}},\
  }\bibfield  {title} {\bibinfo {title} {Local optical imaging of nanoholes
  using a single fluorescent rare--earth-doped glass particle as a probe},\
  }\href@noop {} {\bibfield  {journal} {\bibinfo  {journal} {Applied Physics
  Letters}\ }\textbf {\bibinfo {volume} {83}},\ \bibinfo {pages} {147}
  (\bibinfo {year} {2003})}\BibitemShut {NoStop}%
\bibitem [{\citenamefont {Kasperczyk}\ \emph {et~al.}(2015)\citenamefont
  {Kasperczyk}, \citenamefont {Person}, \citenamefont {Ananias}, \citenamefont
  {Carlos},\ and\ \citenamefont {Novotny}}]{Kasperczyk:15}%
  \BibitemOpen
  \bibfield  {author} {\bibinfo {author} {\bibfnamefont {M.}~\bibnamefont
  {Kasperczyk}}, \bibinfo {author} {\bibfnamefont {S.}~\bibnamefont {Person}},
  \bibinfo {author} {\bibfnamefont {D.}~\bibnamefont {Ananias}}, \bibinfo
  {author} {\bibfnamefont {L.~D.}\ \bibnamefont {Carlos}},\ and\ \bibinfo
  {author} {\bibfnamefont {L.}~\bibnamefont {Novotny}},\ }\bibfield  {title}
  {\bibinfo {title} {Excitation of magnetic dipole transitions at optical
  frequencies},\ }\href@noop {} {\bibfield  {journal} {\bibinfo  {journal}
  {Physical Review Letters}\ }\textbf {\bibinfo {volume} {114}},\ \bibinfo
  {pages} {163903} (\bibinfo {year} {2015})}\BibitemShut {NoStop}%
\bibitem [{\citenamefont {Wiecha}\ \emph {et~al.}(2019)\citenamefont {Wiecha},
  \citenamefont {Majorel}, \citenamefont {Girard}, \citenamefont {Arbouet},
  \citenamefont {Masenelli}, \citenamefont {Boisron}, \citenamefont {Lecestre},
  \citenamefont {Larrieu}, \citenamefont {Paillard},\ and\ \citenamefont
  {Cuche}}]{Wiecha-Cuche:19}%
  \BibitemOpen
  \bibfield  {author} {\bibinfo {author} {\bibfnamefont {P.}~\bibnamefont
  {Wiecha}}, \bibinfo {author} {\bibfnamefont {C.}~\bibnamefont {Majorel}},
  \bibinfo {author} {\bibfnamefont {C.}~\bibnamefont {Girard}}, \bibinfo
  {author} {\bibfnamefont {A.}~\bibnamefont {Arbouet}}, \bibinfo {author}
  {\bibfnamefont {B.}~\bibnamefont {Masenelli}}, \bibinfo {author}
  {\bibfnamefont {O.}~\bibnamefont {Boisron}}, \bibinfo {author} {\bibfnamefont
  {A.}~\bibnamefont {Lecestre}}, \bibinfo {author} {\bibfnamefont
  {G.}~\bibnamefont {Larrieu}}, \bibinfo {author} {\bibfnamefont
  {V.}~\bibnamefont {Paillard}},\ and\ \bibinfo {author} {\bibfnamefont
  {A.}~\bibnamefont {Cuche}},\ }\bibfield  {title} {\bibinfo {title}
  {Enhancement of electric and magnetic dipole transition of rare-earth-doped
  thin films tailored by high-index dielectric nanostructures},\ }\href@noop {}
  {\bibfield  {journal} {\bibinfo  {journal} {Applied Optics}\ }\textbf
  {\bibinfo {volume} {58}},\ \bibinfo {pages} {1682} (\bibinfo {year}
  {2019})}\BibitemShut {NoStop}%
\bibitem [{\citenamefont {Karaveli}\ and\ \citenamefont
  {Zia}(2011)}]{Karaveli-Zia:2011}%
  \BibitemOpen
  \bibfield  {author} {\bibinfo {author} {\bibfnamefont {S.}~\bibnamefont
  {Karaveli}}\ and\ \bibinfo {author} {\bibfnamefont {R.}~\bibnamefont {Zia}},\
  }\bibfield  {title} {\bibinfo {title} {Spectral tuning by selective
  enhancement of electric and magnetic dipole emission},\ }\href@noop {}
  {\bibfield  {journal} {\bibinfo  {journal} {Physical Review Letters}\
  }\textbf {\bibinfo {volume} {106}},\ \bibinfo {pages} {193004} (\bibinfo
  {year} {2011})}\BibitemShut {NoStop}%
\bibitem [{\citenamefont {Aigouy}\ \emph {et~al.}(2014)\citenamefont {Aigouy},
  \citenamefont {Caz\'e}, \citenamefont {Gredin}, \citenamefont {Mortier},\
  and\ \citenamefont {Carminati}}]{Aigouy:14}%
  \BibitemOpen
  \bibfield  {author} {\bibinfo {author} {\bibfnamefont {L.}~\bibnamefont
  {Aigouy}}, \bibinfo {author} {\bibfnamefont {A.}~\bibnamefont {Caz\'e}},
  \bibinfo {author} {\bibfnamefont {P.}~\bibnamefont {Gredin}}, \bibinfo
  {author} {\bibfnamefont {M.}~\bibnamefont {Mortier}},\ and\ \bibinfo {author}
  {\bibfnamefont {R.}~\bibnamefont {Carminati}},\ }\bibfield  {title} {\bibinfo
  {title} {Mapping and quantifying electric and magnetic dipole luminescence at
  the nanoscale},\ }\href@noop {} {\bibfield  {journal} {\bibinfo  {journal}
  {Physical Review Letters}\ }\textbf {\bibinfo {volume} {113}} (\bibinfo
  {year} {2014})}\BibitemShut {NoStop}%
\bibitem [{\citenamefont {Rabouw}\ \emph {et~al.}(2016)\citenamefont {Rabouw},
  \citenamefont {Prins},\ and\ \citenamefont {Norris}}]{Rabouw-Norris:16}%
  \BibitemOpen
  \bibfield  {author} {\bibinfo {author} {\bibfnamefont {F.~T.}\ \bibnamefont
  {Rabouw}}, \bibinfo {author} {\bibfnamefont {P.~T.}\ \bibnamefont {Prins}},\
  and\ \bibinfo {author} {\bibfnamefont {D.~J.}\ \bibnamefont {Norris}},\
  }\bibfield  {title} {\bibinfo {title} {Europium-doped {NaYF$_4$} nanocrystals
  as probes for the electric and magnetic local density of optical states
  throughout the visible spectral range},\ }\href@noop {} {\bibfield  {journal}
  {\bibinfo  {journal} {Nano Letters}\ }\textbf {\bibinfo {volume} {16}},\
  \bibinfo {pages} {7254} (\bibinfo {year} {2016})}\BibitemShut {NoStop}%
\bibitem [{\citenamefont {Li}\ \emph {et~al.}(2018)\citenamefont {Li},
  \citenamefont {Karaveli}, \citenamefont {Cueff}, \citenamefont {Li},\ and\
  \citenamefont {Zia}}]{Li-Zia:18}%
  \BibitemOpen
  \bibfield  {author} {\bibinfo {author} {\bibfnamefont {D.}~\bibnamefont
  {Li}}, \bibinfo {author} {\bibfnamefont {S.}~\bibnamefont {Karaveli}},
  \bibinfo {author} {\bibfnamefont {S.}~\bibnamefont {Cueff}}, \bibinfo
  {author} {\bibfnamefont {W.}~\bibnamefont {Li}},\ and\ \bibinfo {author}
  {\bibfnamefont {R.}~\bibnamefont {Zia}},\ }\bibfield  {title} {\bibinfo
  {title} {Probing the combined electromagnetic local density of optical states
  with quantum emitters supporting strong electric and magnetic transitions},\
  }\href@noop {} {\bibfield  {journal} {\bibinfo  {journal} {Physical Review
  Letters}\ }\textbf {\bibinfo {volume} {121}},\ \bibinfo {pages} {227403}
  (\bibinfo {year} {2018})}\BibitemShut {NoStop}%
\bibitem [{\citenamefont {Binnemans}(2015)}]{Binnemans:15}%
  \BibitemOpen
  \bibfield  {author} {\bibinfo {author} {\bibfnamefont {K.}~\bibnamefont
  {Binnemans}},\ }\bibfield  {title} {\bibinfo {title} {Interpretation of
  europium(III) spectra},\ }\href@noop {} {\bibfield  {journal} {\bibinfo
  {journal} {Coordination Chemistry Reviews}\ }\textbf {\bibinfo {volume}
  {295}},\ \bibinfo {pages} {1} (\bibinfo {year} {2015})}\BibitemShut {NoStop}%
\bibitem [{\citenamefont {Freed}\ and\ \citenamefont
  {Weissman}(1941)}]{Freed-Weissman:1941}%
  \BibitemOpen
  \bibfield  {author} {\bibinfo {author} {\bibfnamefont {S.}~\bibnamefont
  {Freed}}\ and\ \bibinfo {author} {\bibfnamefont {S.}~\bibnamefont
  {Weissman}},\ }\bibfield  {title} {\bibinfo {title} {Multipole nature of
  elementary sources of radiation--wide-angle interference},\ }\href@noop {}
  {\bibfield  {journal} {\bibinfo  {journal} {Physical Review}\ }\textbf
  {\bibinfo {volume} {60}},\ \bibinfo {pages} {440} (\bibinfo {year}
  {1941})}\BibitemShut {NoStop}%
\bibitem [{\citenamefont {Kunz}\ and\ \citenamefont
  {Lukosz}(1980)}]{Kunz-Lukosz:80}%
  \BibitemOpen
  \bibfield  {author} {\bibinfo {author} {\bibfnamefont {R.}~\bibnamefont
  {Kunz}}\ and\ \bibinfo {author} {\bibfnamefont {W.}~\bibnamefont {Lukosz}},\
  }\bibfield  {title} {\bibinfo {title} {Changes in fluorescence lifetimes
  induced by variable optical environments},\ }\href@noop {} {\bibfield
  {journal} {\bibinfo  {journal} {Physical Review B}\ }\textbf {\bibinfo
  {volume} {21}},\ \bibinfo {pages} {4814} (\bibinfo {year}
  {1980})}\BibitemShut {NoStop}%
\bibitem [{\citenamefont {Sayre}\ and\ \citenamefont
  {Freed}(1956)}]{Sayre-Freed:1956}%
  \BibitemOpen
  \bibfield  {author} {\bibinfo {author} {\bibfnamefont {E.~V.}\ \bibnamefont
  {Sayre}}\ and\ \bibinfo {author} {\bibfnamefont {S.}~\bibnamefont {Freed}},\
  }\bibfield  {title} {\bibinfo {title} {Spectra and quantum states of the
  europic ion in crystals. {II.F}luorescence and absorption spectra of single
  crystals of europic ethylsulfate nonahydrate},\ }\href@noop {} {\bibfield
  {journal} {\bibinfo  {journal} {The Journal of Chemical Physics}\ }\textbf
  {\bibinfo {volume} {24}},\ \bibinfo {pages} {1213} (\bibinfo {year}
  {1956})}\BibitemShut {NoStop}%
\bibitem [{\citenamefont {Kim}\ \emph {et~al.}(2017)\citenamefont {Kim},
  \citenamefont {Michelin}, \citenamefont {Hilbers}, \citenamefont
  {Martinelli}, \citenamefont {Chaudan}, \citenamefont {Amselem}, \citenamefont
  {Fradet}, \citenamefont {Boilot}, \citenamefont {Brouwer}, \citenamefont
  {Baroud}, \citenamefont {Peretti},\ and\ \citenamefont
  {Gacoin}}]{Kim-Gacoin:17}%
  \BibitemOpen
  \bibfield  {author} {\bibinfo {author} {\bibfnamefont {J.}~\bibnamefont
  {Kim}}, \bibinfo {author} {\bibfnamefont {S.}~\bibnamefont {Michelin}},
  \bibinfo {author} {\bibfnamefont {M.}~\bibnamefont {Hilbers}}, \bibinfo
  {author} {\bibfnamefont {L.}~\bibnamefont {Martinelli}}, \bibinfo {author}
  {\bibfnamefont {E.}~\bibnamefont {Chaudan}}, \bibinfo {author} {\bibfnamefont
  {G.}~\bibnamefont {Amselem}}, \bibinfo {author} {\bibfnamefont
  {E.}~\bibnamefont {Fradet}}, \bibinfo {author} {\bibfnamefont {J.-P.}\
  \bibnamefont {Boilot}}, \bibinfo {author} {\bibfnamefont {A.~M.}\
  \bibnamefont {Brouwer}}, \bibinfo {author} {\bibfnamefont {C.}~\bibnamefont
  {Baroud}}, \bibinfo {author} {\bibfnamefont {J.}~\bibnamefont {Peretti}},\
  and\ \bibinfo {author} {\bibfnamefont {T.}~\bibnamefont {Gacoin}},\
  }\bibfield  {title} {\bibinfo {title} {Monitoring the orientation of
  rare-earth-doped nanorods for flow shear tomography},\ }\href@noop {}
  {\bibfield  {journal} {\bibinfo  {journal} {Nature Nanotechnology}\ }\textbf
  {\bibinfo {volume} {12}},\ \bibinfo {pages} {914} (\bibinfo {year}
  {2017})}\BibitemShut {NoStop}%
\bibitem [{\citenamefont {Wang}\ \emph {et~al.}(2010)\citenamefont {Wang},
  \citenamefont {Han}, \citenamefont {Lim}, \citenamefont {Lu}, \citenamefont
  {Wang}, \citenamefont {Xu}, \citenamefont {Chen}, \citenamefont {Zhang},
  \citenamefont {Hong},\ and\ \citenamefont {Liu}}]{Wang-Liu:2010}%
  \BibitemOpen
  \bibfield  {author} {\bibinfo {author} {\bibfnamefont {F.}~\bibnamefont
  {Wang}}, \bibinfo {author} {\bibfnamefont {Y.}~\bibnamefont {Han}}, \bibinfo
  {author} {\bibfnamefont {C.~S.}\ \bibnamefont {Lim}}, \bibinfo {author}
  {\bibfnamefont {Y.}~\bibnamefont {Lu}}, \bibinfo {author} {\bibfnamefont
  {J.}~\bibnamefont {Wang}}, \bibinfo {author} {\bibfnamefont {J.}~\bibnamefont
  {Xu}}, \bibinfo {author} {\bibfnamefont {H.}~\bibnamefont {Chen}}, \bibinfo
  {author} {\bibfnamefont {C.}~\bibnamefont {Zhang}}, \bibinfo {author}
  {\bibfnamefont {M.}~\bibnamefont {Hong}},\ and\ \bibinfo {author}
  {\bibfnamefont {X.}~\bibnamefont {Liu}},\ }\bibfield  {title} {\bibinfo
  {title} {Simultaneous phaseand size control of upconversion nanocrystals
  through lanthanide doping},\ }\href@noop {} {\bibfield  {journal} {\bibinfo
  {journal} {Nature}\ }\textbf {\bibinfo {volume} {463}},\ \bibinfo {pages}
  {1061} (\bibinfo {year} {2010})}\BibitemShut {NoStop}%
\bibitem [{\citenamefont {Lem\'enager}\ \emph {et~al.}(2018)\citenamefont
  {Lem\'enager}, \citenamefont {Thiriet}, \citenamefont {Pourcin},
  \citenamefont {Lahlil}, \citenamefont {Valdivia-Valero}, \citenamefont
  {{G. {Colas des Francs}}}, \citenamefont {Gacoin},\ and\
  \citenamefont {Fick}}]{FickOptExp:18}%
  \BibitemOpen
  \bibfield  {author} {\bibinfo {author} {\bibfnamefont {G.}~\bibnamefont
  {Lem\'enager}}, \bibinfo {author} {\bibfnamefont {M.}~\bibnamefont
  {Thiriet}}, \bibinfo {author} {\bibfnamefont {F.}~\bibnamefont {Pourcin}},
  \bibinfo {author} {\bibfnamefont {K.}~\bibnamefont {Lahlil}}, \bibinfo
  {author} {\bibfnamefont {F.}~\bibnamefont {Valdivia-Valero}}, \bibinfo
  {author} {\bibnamefont {{G. {Colas des Francs}}}}, \bibinfo
  {author} {\bibfnamefont {T.}~\bibnamefont {Gacoin}},\ and\ \bibinfo {author}
  {\bibfnamefont {J.}~\bibnamefont {Fick}},\ }\bibfield  {title} {\bibinfo
  {title} {Size-dependent trapping behavior and opticalemission study of
  NaYF$_4$ nanorods in optical fiber tip tweezers},\ }\href@noop {} {\bibfield
  {journal} {\bibinfo  {journal} {Optics Express}\ }\textbf {\bibinfo {volume}
  {26}},\ \bibinfo {pages} {32156} (\bibinfo {year} {2018})}\BibitemShut
  {NoStop}%
\bibitem [{\citenamefont {Walsh}()}]{WalshBM}%
  \BibitemOpen
  \bibfield  {author} {\bibinfo {author} {\bibfnamefont {B.~M.}\ \bibnamefont
  {Walsh}},\ }\bibfield  {title} {\bibinfo {title} {Judd-Ofelt theory:
  Principles and practices},\ }in\ \href@noop {} {\emph {\bibinfo {booktitle}
  {Advances in Spectroscopy for Lasers and Sensing}}},\ \bibinfo {editor}
  {edited by\ \bibinfo {editor} {\bibfnamefont {B.~D.}\ \bibnamefont
  {Bartolo}}\ and\ \bibinfo {editor} {\bibfnamefont {O.}~\bibnamefont
  {Forte}}}\BibitemShut {NoStop}%
\bibitem [{\citenamefont {Tu}\ \emph {et~al.}(2013)\citenamefont {Tu},
  \citenamefont {Liu}, \citenamefont {Zhu}, \citenamefont {Li}, \citenamefont
  {Liu},\ and\ \citenamefont {Chen}}]{Tu-Chen:13}%
  \BibitemOpen
  \bibfield  {author} {\bibinfo {author} {\bibfnamefont {D.}~\bibnamefont
  {Tu}}, \bibinfo {author} {\bibfnamefont {Y.}~\bibnamefont {Liu}}, \bibinfo
  {author} {\bibfnamefont {H.}~\bibnamefont {Zhu}}, \bibinfo {author}
  {\bibfnamefont {R.}~\bibnamefont {Li}}, \bibinfo {author} {\bibfnamefont
  {L.}~\bibnamefont {Liu}},\ and\ \bibinfo {author} {\bibfnamefont
  {X.}~\bibnamefont {Chen}},\ }\bibfield  {title} {\bibinfo {title} {Breakdown
  of crystallographic site symmetry in lanthanide-doped NaYF$_4$ crystals},\
  }\href@noop {} {\bibfield  {journal} {\bibinfo  {journal} {Angewandte
  Communications}\ }\textbf {\bibinfo {volume} {52}},\ \bibinfo {pages} {1128}
  (\bibinfo {year} {2013})}\BibitemShut {NoStop}%
\bibitem [{\citenamefont {Sheppard}\ and\ \citenamefont
  {Matthews}(1987)}]{Sheppard-Matthews:87}%
  \BibitemOpen
  \bibfield  {author} {\bibinfo {author} {\bibfnamefont {C.~J.~R.}\
  \bibnamefont {Sheppard}}\ and\ \bibinfo {author} {\bibfnamefont {H.~J.}\
  \bibnamefont {Matthews}},\ }\bibfield  {title} {\bibinfo {title} {Imaging in
  high-aperture optical systems},\ }\href@noop {} {\bibfield  {journal}
  {\bibinfo  {journal} {J. Opt.Soc. Am. A}\ }\textbf {\bibinfo {volume} {4}},\
  \bibinfo {pages} {1354} (\bibinfo {year} {1987})}\BibitemShut {NoStop}%
\bibitem [{\citenamefont {Drezet}\ \emph {et~al.}(2008)\citenamefont {Drezet},
  \citenamefont {Hohenau}, \citenamefont {Koller}, \citenamefont {Stepanov},
  \citenamefont {Ditlbacher}, \citenamefont {Steinberger}, \citenamefont
  {Aussenegg}, \citenamefont {Leitner},\ and\ \citenamefont
  {Krenn}}]{Drezet2008}%
  \BibitemOpen
  \bibfield  {author} {\bibinfo {author} {\bibfnamefont {A.}~\bibnamefont
  {Drezet}}, \bibinfo {author} {\bibfnamefont {A.}~\bibnamefont {Hohenau}},
  \bibinfo {author} {\bibfnamefont {D.}~\bibnamefont {Koller}}, \bibinfo
  {author} {\bibfnamefont {A.}~\bibnamefont {Stepanov}}, \bibinfo {author}
  {\bibfnamefont {H.}~\bibnamefont {Ditlbacher}}, \bibinfo {author}
  {\bibfnamefont {B.}~\bibnamefont {Steinberger}}, \bibinfo {author}
  {\bibfnamefont {F.}~\bibnamefont {Aussenegg}}, \bibinfo {author}
  {\bibfnamefont {A.}~\bibnamefont {Leitner}},\ and\ \bibinfo {author}
  {\bibfnamefont {J.}~\bibnamefont {Krenn}},\ }\bibfield  {title} {\bibinfo
  {title} {Leakage radiation microscopy of surface plasmon polaritons},\
  }\href@noop {} {\bibfield  {journal} {\bibinfo  {journal} {Materials Science
  and Engineering B}\ }\textbf {\bibinfo {volume} {149}},\ \bibinfo {pages}
  {220} (\bibinfo {year} {2008})}\BibitemShut {NoStop}%
\bibitem [{\citenamefont {Grandidier}\ \emph {et~al.}(2010)\citenamefont
  {Grandidier}, \citenamefont {{Colas des Francs}}, \citenamefont
  {Massenot}, \citenamefont {Bouhelier}, \citenamefont {Weeber}, \citenamefont
  {Markey},\ and\ \citenamefont {Dereux}}]{GrandidierJMicrosc:2010}%
  \BibitemOpen
  \bibfield  {author} {\bibinfo {author} {\bibfnamefont {J.}~\bibnamefont
  {Grandidier}}, \bibinfo {author} {\bibfnamefont {G.}~\bibnamefont
  {{Colas des Francs}}}, \bibinfo {author} {\bibfnamefont
  {S.}~\bibnamefont {Massenot}}, \bibinfo {author} {\bibfnamefont
  {A.}~\bibnamefont {Bouhelier}}, \bibinfo {author} {\bibfnamefont {J.-C.}\
  \bibnamefont {Weeber}}, \bibinfo {author} {\bibfnamefont {L.}~\bibnamefont
  {Markey}},\ and\ \bibinfo {author} {\bibfnamefont {A.}~\bibnamefont
  {Dereux}},\ }\bibfield  {title} {\bibinfo {title} {Leakage radiation
  microscopy of surface plasmon coupled emission: investigation of gain
  assisted propagation in an integrated plasmonic waveguide},\ }\href@noop {}
  {\bibfield  {journal} {\bibinfo  {journal} {Journal of Microscopy}\ }\textbf
  {\bibinfo {volume} {239}},\ \bibinfo {pages} {167} (\bibinfo {year}
  {2010})}\BibitemShut {NoStop}%
\bibitem [{\citenamefont {Lieb}\ \emph {et~al.}(2004)\citenamefont {Lieb},
  \citenamefont {Zavislan},\ and\ \citenamefont {Novotny}}]{lieb04}%
  \BibitemOpen
  \bibfield  {author} {\bibinfo {author} {\bibfnamefont {M.~A.}\ \bibnamefont
  {Lieb}}, \bibinfo {author} {\bibfnamefont {J.}~\bibnamefont {Zavislan}},\
  and\ \bibinfo {author} {\bibfnamefont {L.}~\bibnamefont {Novotny}},\
  }\bibfield  {title} {\bibinfo {title} {Single-molecule orientations
  determined by direct emission pattern imaging},\ }\href@noop {} {\bibfield
  {journal} {\bibinfo  {journal} {J. Opt. Soc. Amer.~B}\ }\textbf {\bibinfo
  {volume} {21}},\ \bibinfo {pages} {1210} (\bibinfo {year}
  {2004})}\BibitemShut {NoStop}%
\bibitem [{\citenamefont {Chang}\ and\ \citenamefont
  {Gruber}(1964)}]{Chang-Gruber:64}%
  \BibitemOpen
  \bibfield  {author} {\bibinfo {author} {\bibfnamefont {N.~C.}\ \bibnamefont
  {Chang}}\ and\ \bibinfo {author} {\bibfnamefont {J.~B.}\ \bibnamefont
  {Gruber}},\ }\bibfield  {title} {\bibinfo {title} {Spectra and energy levels
  of {Eu$^{3+}$} in {Y$_2$O$_3$}},\ }\href@noop {} {\bibfield  {journal}
  {\bibinfo  {journal} {J. Chem. Phys.}\ }\textbf {\bibinfo {volume} {41}},\
  \bibinfo {pages} {3227} (\bibinfo {year} {1964})}\BibitemShut {NoStop}%
\bibitem [{\citenamefont {Rodr\'iguez-Sevilla}\ \emph
  {et~al.}(2016)\citenamefont {Rodr\'iguez-Sevilla}, \citenamefont
  {Labrador-P\`aez}, \citenamefont {nczyk}, \citenamefont {Nyk}, \citenamefont
  {Samo\'c}, \citenamefont {Kar}, \citenamefont {Mackenzie}, \citenamefont
  {Paterson}, \citenamefont {Jaque},\ and\ \citenamefont
  {Haro-Gonz\'alez}}]{RodriguezSevilla:16}%
  \BibitemOpen
  \bibfield  {author} {\bibinfo {author} {\bibfnamefont {P.}~\bibnamefont
  {Rodr\'iguez-Sevilla}}, \bibinfo {author} {\bibfnamefont {L.}~\bibnamefont
  {Labrador-P\`aez}}, \bibinfo {author} {\bibfnamefont {D.~W.}\ \bibnamefont
  {nczyk}}, \bibinfo {author} {\bibfnamefont {M.}~\bibnamefont {Nyk}}, \bibinfo
  {author} {\bibfnamefont {M.}~\bibnamefont {Samo\'c}}, \bibinfo {author}
  {\bibfnamefont {K.}~\bibnamefont {Kar}}, \bibinfo {author} {\bibfnamefont
  {M.~D.}\ \bibnamefont {Mackenzie}}, \bibinfo {author} {\bibfnamefont
  {L.}~\bibnamefont {Paterson}}, \bibinfo {author} {\bibfnamefont
  {D.}~\bibnamefont {Jaque}},\ and\ \bibinfo {author} {\bibfnamefont
  {P.}~\bibnamefont {Haro-Gonz\'alez}},\ }\bibfield  {title} {\bibinfo {title}
  {Determining the 3d orientation of optically trapped upconverting nanorods by
  in situ single-particle polarized spectroscopy},\ }\href@noop {} {\bibfield
  {journal} {\bibinfo  {journal} {Nanoscale}\ }\textbf {\bibinfo {volume}
  {8}},\ \bibinfo {pages} {300} (\bibinfo {year} {2016})}\BibitemShut {NoStop}%
\bibitem [{\citenamefont {Kim}\ \emph {et~al.}(2020)\citenamefont {Kim},
  \citenamefont {Chacon}, \citenamefont {Wang}, \citenamefont {Larquet},
  \citenamefont {Lahlil}, \citenamefont {Leray}, \citenamefont {{G.
  {Colas des Francs}}}, \citenamefont {Kim},\ and\ \citenamefont
  {Gacoin}}]{Kim-Gacoin:20}%
  \BibitemOpen
  \bibfield  {author} {\bibinfo {author} {\bibfnamefont {J.}~\bibnamefont
  {Kim}}, \bibinfo {author} {\bibfnamefont {R.}~\bibnamefont {Chacon}},
  \bibinfo {author} {\bibfnamefont {Z.}~\bibnamefont {Wang}}, \bibinfo {author}
  {\bibfnamefont {E.}~\bibnamefont {Larquet}}, \bibinfo {author} {\bibfnamefont
  {K.}~\bibnamefont {Lahlil}}, \bibinfo {author} {\bibfnamefont
  {A.}~\bibnamefont {Leray}}, \bibinfo {author} {\bibnamefont {{G.
  {Colas des Francs}}}}, \bibinfo {author} {\bibfnamefont {J.}~\bibnamefont
  {Kim}},\ and\ \bibinfo {author} {\bibfnamefont {T.}~\bibnamefont {Gacoin}},\
  }\bibfield  {title} {\bibinfo {title} {Measuring 3d orientation of
  nanocrystals via polarized luminescence of rare-earth dopants},\ }\href@noop
  {} {\bibfield  {journal} {\bibinfo  {journal} {submitted}\ } (\bibinfo {year}
  {2020})}\BibitemShut {NoStop}%
\bibitem [{\citenamefont {Novotny}\ and\ \citenamefont
  {Hecht}(2006)}]{Novotny-Hecht:2006}%
  \BibitemOpen
  \bibfield  {author} {\bibinfo {author} {\bibfnamefont {L.}~\bibnamefont
  {Novotny}}\ and\ \bibinfo {author} {\bibfnamefont {B.}~\bibnamefont
  {Hecht}},\ }\href@noop {} {\emph {\bibinfo {title} {Principles of
  Nano-Optics}}},\ edited by\ \bibinfo {editor} {\bibfnamefont {C.~U.}\
  \bibnamefont {Press}}\ (\bibinfo  {publisher} {Cambridge University Press},\
  \bibinfo {year} {2006})\BibitemShut {NoStop}%
\end{thebibliography}
%

\end{document}